\documentstyle[prd,preprint,tighten,aps,eqsecnum,%
amssymb,newlfont,epsfig,floats]{revtex}
\begin{document}
\preprint{
\begin{tabular}{r}
UWThPh-1999-63\\
September 2000
\end{tabular}
}

\vspace{2cm}

\title{Two-dimensional gravitational anomalies, Schwinger terms and dispersion relations\thanks{This work
was partly supported by Austria-Czech Republic Scientific collaboration, project KONTACT
1999-8.}}

\vspace{2cm}

\author{R.A. Bertlmann and E. Kohlprath\thanks{Supported by a Wissenschaftsstipendium der
Magistratsabteilung 18 der Stadt Wien.}}
\address{Institut f\"ur Theoretische Physik,
Universit\"at Wien\\
Boltzmanngasse 5,
A-1090 Vienna, Austria}

\maketitle

\vspace{2cm}

\begin{abstract}
We are dealing with two-dimensional
gravitational anomalies, specifically with the Einstein anomaly and
the Weyl anomaly, and we show that they are fully determined by dispersion relations independent of any
renormalization procedure (or ultraviolet regularization).
The origin of the anomalies is the existence of a superconvergence sum rule for the
imaginary part of the relevant formfactor. In the zero mass limit the imaginary part
of the formfactor approaches a $\delta$-function singularity at zero momentum squared,
exhibiting in this way the infrared feature of the gravitational anomalies.
We find an equivalence between the dispersive approach and the dimensional
regularization procedure. The Schwinger terms appearing in the equal time commutators
of the energy momentum tensors can be calculated by the same dispersive method.
Although all computations are performed in two dimensions the method is expected to work in higher dimensions too.

\end{abstract}

\newpage

\section{Introduction}

An anomaly in field theory occurs if a symmetry of the action or the corresponding
conservation law, valid in the classical theory, is violated in the quantized version.
This surprising feature of quantum theory discovered by Adler \cite{Adler}, Bell and
Jackiw \cite{BellJackiw}, and by Bardeen \cite{Bardeen} in 1969 plays a fundamental role
in physics (for details see Refs.\cite{Bertlmann} -- \cite{Jackiw2}).
Physically there is a difference between external and internal symmetries. The
breakdown of an external symmetry is not dangerous for the consistency of the
theory, on the contrary, it provides for instance the physical explanation for the
$\pi^{0}$-decay \cite{Adler}, \cite{BellJackiw}
or the solution to the $U(1)$ problem in QCD \cite{Hooft}.
On the other hand, the breakdown of an internal symmetry (i.e. gauge symmetry) leads to an
inconsistency of the quantum theory, the anomalous Ward identities destroy the
renormalizability of the theory \cite{GrossJackiw}, and also the unitarity of the
$S$-matrix may be lost \cite{KorthalsAltesPerrottet}.
To avoid such anomalies imposes severe restrictions to the physical content of a
theory. For instance, in the famous $SU(2) \times U(1)$ standard theory for electroweak
interactions one had to demand the existence of the top quark long before it was discovered.

Gravitation regarded as a gauge theory also suffers from anomalies. The gauges are the
general coordinate transformations (diffeomorphisms) or the rotations in the tangent frame
(Lorentz transformations) or the conformal transformations (Weyl transformations). Then
in the quantum case the classical conservation law of the energy-momentum tensor can be
broken -- an Einstein anomaly occurs -- or an antisymmetric part of the energy-momentum
tensor can exist -- a Lorentz anomaly occurs -- or the trace of the tensor is nonvanishing
-- a Weyl anomaly arises (for details see e.g. Refs.\cite{Bertlmann},\cite{Kohlprath},
\cite{AlvarezGinspargAnn}).
Whereas the anomaly in the tensor trace has been found already in the seventies 
\cite{CapperDuffNC} -- \cite{DeserNP}
the study
of the gravitational anomalies started with the pioneering work of Alvarez-Gaum\'e and
Witten \cite{AlvarezWitten} in the eighties. The anomalies have been first found within
perturbation theory, they are local polynomials in the connection and curvature. The authors
\cite{AlvarezWitten} -- \cite{Langouche}
have calculated (ultraviolet divergent) Feynman diagrams where the external gravitational
field couples to a fermion loop via the energy-momentum tensor. Of course, the anomaly
-- reflecting the deep laws of quantum physics -- must show up within other approaches too.
So they have been calculated by the heat kernel method \cite{Leutwyler},
\cite{LeutwylerMallik}, by Fujikawa's path integral approach \cite{Fujikawa},
\cite{FujikawaTomiyaYasuda}, and by modern mathematical techniques such as differential
geometry and cohomology \cite{Stora84} -- \cite{BonoraPastiTonin} and topology
(index theorems) \cite{AlvarezGinspargAnn}, \cite{AlvarezSingerZumino}, \cite{Richter},
(for an overview see Ref.\cite{Bertlmann}).  

Deeply related to anomalies are the socalled Schwinger terms
\cite{Schwinger51} -- \cite{GotoImamura} (for an introduction see e.g.
Refs. \cite{Jackiw1}, \cite{Jackiw2}). In a Yang-Mills gauge theory
Schwinger terms (ST) show up as additional terms (extensions) in the canonical
algebra of the equal time commutators (ETC) of the Gauss law operators (see e.g.
\cite{Pawlowski98} -- \cite{Sykora}).
Cohomologically they are described by the Faddeev-Mickelsson cocycle \cite{Faddeev},
\cite{Mickelsson}, and geometrically they can be related to a Berry phase in the
vacuum functional \cite{NiemiSemenoff} -- \cite{Rupp}.
ST are frequently calculated within perturbation theory where the Bjorken-Johnson-Low limit
\cite{Bjorken}, \cite{JohnsonLow} works very well.
However, the definition of a point-splitting method turns out to be more subtle and might
not lead to the correct result \cite{BertlmannSykora}, \cite{Jo}.

In gravitation Schwinger terms occur in the ETC of the energy-momentum tensors, c-number
terms that are proportional to derivatives of the $\delta$-function. They can be related
to the gravitational anomalies \cite{EbnerHeidLopes}, and they have been calculated
explictly via the invariant spectral function and via cohomological techniques \cite{Tomiya}.
Furthermore there exists an interesting relation of the ST to the curvature of the
determinant line bundle \cite{EkstrandMickelsson}, \cite{Ekstrand}.

Our work deals with the calculation of the gravitational anomalies, specifically the
Einstein anomaly and the Weyl anomaly. The Lorentz anomaly is not 
independent of the Einstein anomaly, both types of anomalies can be shifted into each other
by a suitable counterterm \cite{BardeenZumino}. For convenience we choose the case where the
Lorentz anomaly is vanishing. We also calculate the gravitational Schwinger terms. 
The purpose of our work is to show that all these anomalous features are easily obtained by the
method of dispersion relations, a less familiar but very useful approach. Some of our results
we have briefly presented in Refs. \cite{BertlmannKohlprathGR}, \cite{BertlmannKohlprathST}.

Already since their first introduction into quantum field theory
\cite{GellMannThirringGoldberger} dispersion relations (DR) proved to be a very valuable tool.
In connection with anomalies DR have been fomulated
by Dolgov and Zakharov \cite{DolgovZakharov} and also by Kummer \cite{Kummer}.
In the following several authors \cite{HorejsiPRD} -- \cite{AdamBertlmannHofer1} used 
successfully DR to determine the anomalies in the chiral current.
Recently Ho\v rej\v s\' \i{} and Schnabl \cite{HorejsiSchnabl} have applied the method to
the well-known trace anomaly \cite{ChanowitzEllis}, \cite{AdlerCollinsDuncan} which is
related to the broken dilatation (or scale) invariance. We extend in our work the method of
DR to the case of pure gravitation. So we consider chiral fermion loops coupled to
gravitation -- for their evaluation it is enough to use gravitation as an
external or nonquantized field -- and we show that the gravitational anomalies and the
Schwinger terms are, in fact, completely determined by dispersion relations.
All calculations are performed in two dimensions.

Conceptuelly the DR approach is an independent and complementary view of the anomaly
phenomenon as compared to the ultraviolet regularization procedures. Within DR the anomaly
manifests itself as a very peculiar infrared feature of the imaginary part of the amplitude.
But as we shall show there is a link between the two approaches, the DR method and the
n-dimensional regularization precedure.

Our paper is organized as follows. In Section 2 we present the general structure of the
considered (pseudo-) tensor amplitude and we discuss the Ward identities which we have to
study. In Section 3 we introduce the dispersion relations for the relevant formfactors of
the amplitude and calculate their imaginary parts via the Cutkosky rule. In order to
reproduce the DR results in a definite ultraviolet regularization scheme we have worked out
in detail the 't~Hooft-Veltman regularization procedure in Section 4 and we have compared
the several amplitudes with the results of Tomiya \cite{Tomiya}
and Alvarez-Gaum\'e  and Witten \cite{AlvarezWitten}.
The equivalence between the dispersive approach and the dimensional regularization procedure
is given in Section 5. In Section 6 we derive the anomalous Ward identities and explain the
source of the anomaly in the DR approach. From the Ward identities we deduce the linearized
gravitational anomalies -- the Einstein- and the Weyl anomaly -- and we also determine their
covariant versions, a comparison with the exact results is given.
The gravitational Schwinger terms occuring in the ETC of the energy-momentum
tensors we calculate in Section 7, where we adapt the dispersive approach to a method
proposed by K\"all\'en \cite{Kaellen}. Finally we summarize our main results in Section 8.

\newpage

\section{Structure of the amplitude}

In two dimensions the Lagrangian describing a Weyl fermion in a gravitational
background field can be written as

\begin{equation} \label{Lagrangian}
\mathcal{L} = i e E^{a\mu} \bar \psi \gamma_{a} \frac{1}{2} \stackrel{\leftrightarrow}{D}_{\mu} \frac{1\pm\gamma_{5}}{2} \psi \, ,
\end{equation}
where $e^{a}_{\,\,\mu}$ is the zweibein and $E_{a}^{\,\,\mu}$ its inverse
$E_{a}^{\,\,\mu}e^{a}_{\,\,\nu} = \delta^{\mu}_{\,\,\nu}$. The determinant of the zweibein is
$e = |det\, e^{a}_{\,\,\mu}|$ and $D_{\mu} = \partial_{\mu} + \omega_{\mu}$ is the covariant
derivative with the spin connection $\omega_{\mu}$.

We use the following conventions in 2 dimensions:

\begin{itemize}
\item for the flat metric
\begin{equation} \label{definition of the flat metric}
\eta_{\mu\nu} = \left( \begin{array}{cc}1&\\&-1 \end{array} \right)
\end{equation}
\item for the epsilon tensor
\begin{equation} \label{definition of the epsilon tensor}
\varepsilon^{\mu\nu} = \left( \begin{array}{cc} 0&1\\-1&0 \end{array} \right)
\end{equation}
\item for the Dirac matrices
\begin{eqnarray} 
\gamma^{0}&=&\sigma^{2}= \left(\begin{array}{cc}0&-i\\i&0\end{array}\right)  \\
\gamma^{1}&=&i\sigma^{1}=\left(\begin{array}{cc}0&i\\i&0\end{array}\right) \\
\gamma_{5}&=&\gamma^{0}\gamma^{1}=\sigma^{3}=\left(\begin{array}{cc}1&0\\0&-1\end{array}\right).\label{definition of the gamma matrices}
\end{eqnarray}
\end{itemize}

\begin{figure}
\centering
\resizebox{10cm}{!}{\includegraphics{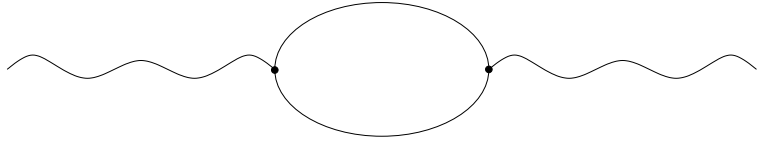}}
\caption{Weyl fermion contribution to the graviton propagator}
\label{Loop}
\end{figure}

The Einstein and the Weyl anomaly are determined by the one-loop diagram in 
Fig.\ref{Loop} and it is sufficient to use the linearized gravitational field

\begin{eqnarray}
&&g_{\mu\nu}=\eta_{\mu\nu}+\kappa h_{\mu\nu}+O(\kappa^{2}) \, ,\quad g^{\mu\nu}=\eta^{\mu\nu}-\kappa h^{\mu\nu}+O(\kappa^{2}) \, ,\\
&&e^{a}_{\ \mu}=\eta^{a}_{\ \mu}+\frac{1}{2}\kappa h^{a}_{\ \mu}+O(\kappa^{2}) \, ,\quad  E^{a\mu} = \eta^{a\mu} - \frac{1}{2}\kappa h^{a\mu}+O(\kappa^{2}) \, .\label{linearization of the vielbein with upper indices}
\end{eqnarray}

Since in two dimensions the spin connection $\omega_{\mu}$ does not contribute
(see e.g. Ref.\cite{Bertlmann}) we find the following linearized interaction Lagrangian 
(for convenience $\kappa$ is absorbed into $h^{a\mu}$, $\partial_{\mu}^{\psi}$ acts only
on $\psi$)

\begin{equation}
\mathcal{L}^{lin}_{I} = - \frac{i}{4} \left( h^{a\mu} \bar\psi \gamma_{a} \frac{1\pm\gamma_{5}}{2} \stackrel{\leftrightarrow}{\partial^{\psi}_{\mu}} \psi + h^{\mu}_{\ \mu} \bar\psi \gamma^{a} \frac{1\pm\gamma_{5}}{2} \stackrel{\leftrightarrow}{\partial^{\psi}_{a}} \psi \right)
\end{equation}
and
\begin{equation} \label{linearized interaction Lagrangian}
\mathcal{L}^{lin}_{I} = -\frac{1}{2}  h_{\mu\nu} T^{\mu\nu} \, .
\end{equation}
From this expression follow the Feynman rules for the vertices in the loop diagram

\begin{equation}
- \frac{i}{4} \left( \gamma_{\mu} (k_{1} - k_{2})_{\nu} + \gamma_{\nu} (k_{1} - k_{2})_{\mu} \right) \frac{1\pm\gamma_{5}}{2}
\end{equation}
and the explicit form of the (symmetric) energy-momentum tensor

\begin{eqnarray} \label{energy-momentum-tensor}
T^{\mu\nu} &=& \frac{1}{2} \left( T^{\mu}_{\ a} E^{a\nu} + T^{\nu}_{\ a} E^{a\mu} \right) {}\nonumber\\{} &=& \frac{i}{4} \Biggl( \bar\psi E^{a\nu} \gamma_{a} \frac{1\pm\gamma_{5}}{2} \stackrel{\leftrightarrow}{\ D^{\mu}} \psi + \bar\psi E^{a\mu} \gamma_{a} \frac{1\pm\gamma_{5}}{2}  \stackrel{\leftrightarrow}{\ D^{\nu}} \psi \Biggr),
\end{eqnarray}
where we have dropped the terms proportional to $g_{\mu\nu}$ as they do not contribute
to the amplitude.\\

Then the whole amplitude is given by the two-point function

\begin{equation} \label{amplitude expressed by T-product}
T_{\mu \nu \rho \sigma}(p) = i \int d^{2}x e^{i p x} \langle 0 \vert T \lbrack T_{\mu\nu}(x) T_{\rho\sigma}(0) \rbrack \vert 0 \rangle.
\end{equation}
Due to Lorentz covariance and symmetry the general structure of the amplitude can be
written in the following way

\begin{equation}
T_{\mu \nu \rho \sigma}=T^{V}_{\mu \nu \rho \sigma}+T^{A}_{\mu \nu \rho \sigma}\label{formfactors1}
\end{equation}

\begin{eqnarray}
T^{V}_{\mu \nu \rho \sigma}(p) &=& p_{\mu}p_{\nu}p_{\rho}p_{\sigma} T_{1}(p^{2}) + (p_{\mu}p_{\nu} g_{\rho \sigma} + p_{\rho}p_{\sigma} g_{\mu \nu}) T_{2}(p^{2}) {} \nonumber \\ {} & &+ (p_{\mu}p_{\rho} g_{\nu \sigma} + p_{\mu}p_{\sigma} g_{\nu \rho} + p_{\nu}p_{\rho} g_{\mu \sigma} + p_{\nu}p_{\sigma} g_{\mu \rho}) T_{3}(p^{2}) {} \nonumber \\ {} & &+ g_{\mu \nu}g_{\rho \sigma} T_{4}(p^{2}) + (g_{\mu \rho}g_{\nu \sigma} + g_{\mu \sigma}g_{\nu \rho}) T_{5}(p^{2})\label{formfactors2}
\end{eqnarray}

\begin{eqnarray}
T^{A}_{\mu \nu \rho \sigma}(p) &=& (\varepsilon_{\mu \tau} p^{\tau} p_{\nu} p_{\rho} p_{\sigma} + \varepsilon_{\nu \tau} p^{\tau} p_{\mu} p_{\rho} p_{\sigma} + \varepsilon_{\rho \tau} p^{\tau} p_{\mu} p_{\nu} p_{\sigma} + \varepsilon_{\sigma \tau} p^{\tau} p_{\mu} p_{\nu} p_{\rho}) T_{6}(p^{2}) {} \nonumber \\ {}& & + (\varepsilon_{\mu \tau} p^{\tau} p_{\nu} g_{\rho \sigma} + \varepsilon_{\nu \tau} p^{\tau} p_{\mu} g_{\rho \sigma} + \varepsilon_{\rho \tau} p^{\tau} p_{\sigma} g_{\mu \nu} + \varepsilon_{\sigma \tau} p^{\tau} p_{\rho} g_{\mu \nu}) T_{7}(p^{2}) {} \nonumber \\ {}& &+ \Bigl[\varepsilon_{\mu \tau} p^{\tau} (p_{\rho} g_{\nu \sigma} + p_{\sigma} g_{\nu \rho}) + \varepsilon_{\nu \tau} p^{\tau} (p_{\rho} g_{\mu \sigma} + p_{\sigma} g_{\mu \rho})  {} \nonumber \\ {}& & +\varepsilon_{\rho \tau} p^{\tau} (p_{\mu} g_{\nu \sigma} + p_{\nu} g_{\mu \sigma}) + \varepsilon_{\sigma \tau} p^{\tau} (p_{\mu} g_{\nu \rho} + p_{\nu} g_{\mu \rho})\Bigr] T_{8}(p^{2}) \, ,\label{formfactors3}
\end{eqnarray}
where we have separated the amplitude into its pure tensor part
$T^{V}_{\mu\nu\rho\sigma}$ (coming from the vector piece of the chirality projection
in Eq.(\ref{energy-momentum-tensor})) and into its pseudo-tensor part
$T^{A}_{\mu\nu\rho\sigma}$ (coming from the axial piece in
Eq.(\ref{energy-momentum-tensor})). The functions $T_{1}(p^{2}), ... ,T_{8}(p^{2})$
are the formfactors that are to be evaluated.\\

Classically the energy-momentum tensor has the following properties:
\begin{enumerate}
\item $T_{\mu\nu} = T_{\nu\mu}$, symmetric
\item $\nabla^{\mu}T_{\mu\nu} = 0$, conserved
\item $T^{\mu}_{\ \mu} = 0$, traceless
\end{enumerate}
which lead to the canonical (naive) Ward identities:
\begin{enumerate}
\item $T_{\mu\nu\rho\sigma}(p) = T_{\nu\mu\rho\sigma}(p)$
\item $p^{\mu}T_{\mu\nu\rho\sigma}(p) = 0$
\item $g^{\mu\nu}T_{\mu\nu\rho\sigma}(p) = 0$\,.
\end{enumerate}

We are interested in the pure Einstein anomaly therefore we demand the quantized
energy-momentum tensor to be symmetric. This is always possible due to the
Bardeen--Zumino theorem \cite{BardeenZumino} which states that the gravitational
anomaly can be shifted from pure Lorentz type to pure Einstein type, and vice versa.
(We disregard here Leutwyler's point of view \cite{Leutwyler,LeutwylerMallik} who
emphasizes his preference for the Lorentz anomaly). Thus the symmetry property 1.) of
the amplitude is fulfilled and an other symmetry
$T_{\mu\nu\rho\sigma}(p) = T_{\rho\sigma\mu\nu}(p)$ is trivially satisfied. However,
the naive Ward identities 2.) and 3.) need not be satisfied, they can be broken by
the Einstein and Weyl anomalies respectively.

The canonical Ward identities we re-express by the formfactors

\begin{eqnarray}
p^{\mu} T^{V}_{\mu \nu \rho \sigma}(p) &=& p_{\nu}p_{\rho}p_{\sigma}(p^{2} T_{1} + T_{2} + 2 T_{3}) + p_{\nu} g_{\rho \sigma}(p^{2} T_{2} + T_{4}) {}\nonumber\\{}&&+ (p_{\rho} g_{\nu \sigma} + p_{\sigma} g_{\nu \rho})(p^{2} T_{3} + T_{5})  \label{pTV}\\
p^{\mu} T^{A}_{\mu \nu \rho \sigma}(p) &=& \varepsilon_{\nu \tau} p^{\tau}\left[p_{\rho}p_{\sigma} (p^{2} T_{6} + 2 T_{8}) + g_{\rho \sigma} p^{2}T_{7}\right] {}\nonumber\\{}&&+ \varepsilon_{\rho \tau} p^{\tau}\left[p_{\nu}p_{\sigma} (p^{2} T_{6} + T_{7} + T_{8}) + g_{\nu \sigma} p^{2} T_{8}\right] {}\nonumber\\{}&&+  \varepsilon_{\sigma \tau} p^{\tau}\left[p_{\nu}p_{\rho} (p^{2} T_{6} + T_{7} + T_{8}) + g_{\nu \rho} p^{2} T_{8}\right] \label{pTA}
\end{eqnarray}
and
\begin{eqnarray}
g^{\mu\nu} T^{V}_{\mu \nu \rho \sigma}(p) &=& p_{\rho}p_{\sigma} (p^{2} T_{1} + 2\omega T_{2} + 4 T_{3}) + g_{\rho\sigma} (p^{2} T_{2} + 2\omega T_{4} + 2 T_{5}) \label{gTV}\\
g^{\mu\nu} T^{A}_{\mu \nu \rho \sigma}(p) &=& (\varepsilon_{\rho\tau} p^{\tau}p_{\sigma} + \varepsilon_{\sigma\tau} p^{\tau}p_{\rho}) (p^{2} T_{6} + 2\omega T_{7} + 4 T_{8}) \, , \label{gTA}
\end{eqnarray} 
where $n=2\omega$ is the dimension. We call from now on Ward identity (WI) the
property 2.) and trace identity (TI) the property 3.). For the pure tensor part of the
amplitude the WI may be written as

\begin{eqnarray}
&&p^{2} T_{1} + T_{2} + 2 T_{3} = 0 \label{VWI1} \\
&&p^{2} T_{2} + T_{4} = 0 \label{VWI2} \\
&&p^{2} T_{3} + T_{5} = 0 \, , \label{VWI3}
\end{eqnarray}
and the TI as

\begin{eqnarray}
&&p^{2} T_{1} + 2\omega T_{2} + 4 T_{3} = 0 \label{TI1} \\
&&p^{2} T_{2} + 2\omega T_{4} + 2 T_{5} = 0 \, . \label{TI2}
\end{eqnarray}
Of course, relations (\ref{TI2}) and (\ref{TI1}) are not independent of each other, 
(\ref{TI2}) follows from (\ref{TI1}) and (\ref{VWI1}) -- (\ref{VWI3}).

In the following we shall use a renormalization procedure which keeps the WI in the pure
tensor part (\ref{VWI1}) -- (\ref{VWI3}) so that the anomaly occurs only in the axial
pieces -- representing the pseudotensor part of the amplitude.\\

Calculating the amplitude with massive fermions (the loop in Fig.~\ref{Loop}) with help of
the Feynman rules gives

\begin{eqnarray} \label{amplitude by Dimensional regularization with massive fermions}
T_{\mu \nu \rho \sigma}(p)&=& -\frac{i}{16} Tr \int\!\! \frac{d^{2\omega}k}{(2\pi)^{2\omega}} \Biggl\{\left[\gamma_{\mu} (p+2k)_{\nu} + \gamma_{\nu} (p+2k)_{\mu}\right] \frac{1\pm\gamma_{5}}{2}{}\nonumber\\&&{}\times \frac{/\hspace{-6.5pt}p + /\hspace{-6.5pt}k+m}{(p+k)^{2}-m^{2}+i \varepsilon} \left[ \gamma_{\rho} (p+2k)_{\sigma} + \gamma_{\sigma} (p+2k)_{\rho}\right] \frac{1\pm\gamma_{5}}{2} \frac{/\hspace{-6pt}k+m}{k^{2}-m^{2}+i \varepsilon} \Biggr\}.
\end{eqnarray}
For convenience we split the pure tensor piece of the loop in the following way

\begin{equation} \label{pure vector amplitude}
T^{V}_{\mu \nu \rho \sigma}(p) = \frac{1}{2} T^{pv}_{\mu \nu \rho \sigma}(p) - T^{dv}_{\mu \nu \rho \sigma}(p) \, ,
\end{equation}
where $T^{pv}_{\mu\nu\rho\sigma}$ represents the loop with the identity instead of the
chirality projectors and $T^{dv}_{\mu\nu\rho\sigma}$ denotes the part proportional
to $m^{2}$. We separate the 'no interchange' amplitudes as

\begin{equation} \label{without interchange}
T^{V}_{\mu \nu \rho \sigma} = T^{ni}_{\mu \nu \rho \sigma} + T^{ni}_{\nu \mu \rho \sigma} + T^{ni}_{\mu \nu \sigma \rho} + T^{ni}_{\nu \mu \sigma \rho} \, .
\end{equation}
Finally, the axial part of the amplitude is connected to the vector part due to
relation

\begin{equation} \label{elimination of gamma5}
\gamma_{\mu}\gamma_{5} = -\varepsilon_{\mu \nu} \gamma^{\nu}
\end{equation}
(valid only in 2 dimensions for our conventions (\ref{definition of the flat metric}) - (\ref{definition of the gamma matrices})). A symmetric decomposition turns out to be useful
\begin{eqnarray} \label{axial part with gamma5 symmetrically distributed}
T^{A}_{\mu \nu \rho \sigma}(p) &=& \mp \Biggl\{ \frac{1}{2} (\varepsilon_{\mu}^{\ \tau} T^{ni}_{\tau \nu \rho \sigma} + \varepsilon_{\rho}^{\ \tau} T^{ni}_{\mu \nu \tau \sigma}) + \frac{1}{2} (\varepsilon_{\nu}^{\ \tau} T^{ni}_{\tau \mu \rho \sigma} + \varepsilon_{\rho}^{\ \tau} T^{ni}_{\nu \mu \tau \sigma}) {} \nonumber \\ {}& &+ \frac{1}{2} (\varepsilon_{\mu}^{\ \tau} T^{ni}_{\tau \nu \sigma \rho} + \varepsilon_{\sigma}^{\ \tau} T^{ni}_{\mu \nu \tau \rho}) + \frac{1}{2} (\varepsilon_{\nu}^{\ \tau} T^{ni}_{\tau \mu \sigma \rho} + \varepsilon_{\sigma}^{\ \tau} T^{ni}_{\nu \mu \tau \rho}) \Biggr\}.
\end{eqnarray}

\section{Dispersion relations}

The formfactors of the amplitude (\ref{formfactors1}) -- (\ref{formfactors3}) are
analytic functions in the complex
$p^{2} = t$ plane except a cut on the real axis starting at $t = 4m^{2}$. Due to
Cauchy's theorem they can be expressed by dispersion relations which relate the
real part of the amplitude to its imaginary part. 

The imaginary parts of the amplitude can be easily calculated via Cutkosky's rule
\cite{Cutkosky}. It states to replace in the amplitude each propagator by its
discontinuity on mass shell

\begin{eqnarray}
Im T^{ni}_{\mu \nu \rho \sigma}(p) &=& \frac{1}{32} \int \!d^{2}k (p+2k)_{\nu} (p+2k)_{\sigma}{}\nonumber\\{}&&\!\!\!\times \Biggl[ (p+k)_{\mu}k_{\rho} + (p+k)_{\rho}k_{\mu} - g_{\mu\rho} (p+k)^{\lambda}k_{\lambda} \Biggr] {}\nonumber\\{}&& \!\!\!\times \delta(k^{2}-m^{2}) \delta\Bigl((p+k)^{2}-m^{2}\Bigr) \theta(-k^{0}) \theta(k^{0}+p^{0})  \label{abstract Im Tni}\\
Im T^{dv,ni}_{\mu \nu \rho \sigma}(p) &=& \frac{1}{32} m^{2} g_{\mu\rho} \int \!d^{2}k (p+2k)_{\nu} (p+2k)_{\sigma}{}\nonumber\\{}&& \!\!\!\times \delta(k^{2}-m^{2}) \delta\Bigl((p+k)^{2}-m^{2}\Bigr) \theta(-k^{0}) \theta(k^{0}+p^{0}) \, . \label{abstract Im Tdvni}
\end{eqnarray}
The explicit integration gives

\begin{eqnarray}
&&Im T^{ni}_{\mu \nu \rho \sigma}(p) = \frac{1}{32} J_{0} \Biggl\{ \left[ -2\frac{m^{2}}{p^{2}}+8\frac{m^{4}}{(p^{2})^{2}} \right] p_{\mu}p_{\nu}p_{\rho}p_{\sigma}{}\nonumber\\{}&&  \qquad + \left[ -\frac{1}{6}p^{2}+\frac{4}{3}m^{2}-\frac{8}{3}\frac{m^{4}}{p^{2}}\right] \left(p_{\mu}p_{\nu} g_{\rho \sigma} + p_{\mu}p_{\sigma} g_{\nu \rho} + p_{\nu}p_{\rho} g_{\mu \sigma} + p_{\rho}p_{\sigma} g_{\mu \nu}\right)  {}\nonumber\\{}&& \qquad + \left[ \frac{1}{3}p^{2}-\frac{2}{3}m^{2}-\frac{8}{3}\frac{m^{4}}{p^{2}}\right] p_{\mu}p_{\rho} g_{\nu \sigma} + \left[\frac{1}{3}p^{2}-\frac{5}{3}m^{2}+\frac{4}{3}\frac{m^{4}}{p^{2}}\right] p_{\nu}p_{\sigma} g_{\mu \rho} {}\nonumber\\{}&&  \qquad + \left[\frac{1}{6}(p^{2})^{2}-\frac{4}{3}p^{2}m^{2}+\frac{8}{3}m^{4}\right] \left(g_{\mu \nu} g_{\rho \sigma} + g_{\mu \sigma} g_{\nu \rho}\right) {}\nonumber\\{}&&\qquad + \left[-\frac{1}{3}(p^{2})^{2}+\frac{5}{3}p^{2}m^{2}-\frac{4}{3}m^{4}\right] g_{\mu \rho} g_{\nu \sigma} \Biggr\} \\
&&Im T^{dv,ni}_{\mu \nu \rho \sigma}(p) = \frac{1}{32} J_{0} m^{2} g_{\mu\rho} \Biggl\{ \left( -p^{2}+4m^{2} \right) g_{\nu\sigma} + \left(1-4\frac{m^{2}}{p^{2}} \right) p_{\nu}p_{\sigma} \Biggr\},
\end{eqnarray}
with the threshold function
\begin{equation}
J_{0} = \frac{1}{p^{2}}\left(1-\frac{4m^{2}}{p^{2}}\right)^{-1/2}\theta(p^{2}-4m^{2}) \, ,
\end{equation}
from which we quickly find all imaginary parts of the formfactors in the total
amplitude $T_{\mu\nu\rho\sigma}$:

\begin{eqnarray}
Im T_{1}(p^{2}) &=& -\frac{1}{4} J_{0}\frac{m^{2}}{p^{2}}\left(1-4\frac{m^{2}}{p^{2}}\right) \\
Im T_{2}(p^{2}) &=& -\frac{1}{48} J_{0}\,p^{2}\left(1-8\frac{m^{2}}{p^{2}}+16\frac{m^{4}}{p^{4}}\right) \\
Im T_{3}(p^{2}) &=& \frac{1}{96} J_{0}\,p^{2}\left(1+\frac{m^{2}}{p^{2}}-20\frac{m^{4}}{p^{4}}\right) \\
Im T_{4}(p^{2}) &=& \frac{1}{48} J_{0}\,p^{4}\left(1-8\frac{m^{2}}{p^{2}}+16\frac{m^{4}}{p^{4}}\right) \\
Im T_{5}(p^{2}) &=& -\frac{1}{96} J_{0}\,p^{4}\left(1-2\frac{m^{2}}{p^{2}}-8\frac{m^{4}}{p^{4}}\right) \\
Im T_{6}(p^{2}) &=& \pm \frac{1}{16} J_{0}\frac{m^{2}}{p^{2}}\left(1-4\frac{m^{2}}{p^{2}}\right) \\
Im T_{7}(p^{2}) &=& \pm \frac{1}{192} J_{0}\,p^{2}\left(1-8\frac{m^{2}}{p^{2}}+16\frac{m^{4}}{p^{4}}\right) \\
Im T_{8}(p^{2}) &=& \mp \frac{1}{384} J_{0}\,p^{2}\left(1+4\frac{m^{2}}{p^{2}}-32\frac{m^{4}}{p^{4}}\right).
\end{eqnarray}

Considering in addition the amplitude $T^{pv}_{\mu\nu\rho\sigma}$
we have the following imaginary parts of the formfactors which we denote by:

\begin{eqnarray}
Im A_{1}(p^{2}) &=& -\frac{1}{2} J_{0}\frac{m^{2}}{p^{2}}\left(1-4\frac{m^{2}}{p^{2}}\right) \label{Im A1}\\
Im A_{2}(p^{2}) &=& -\frac{1}{24} J_{0}\,p^{2}\left(1-8\frac{m^{2}}{p^{2}}+16\frac{m^{4}}{p^{4}}\right) \\
Im A_{3}(p^{2}) &=& \frac{1}{48} J_{0}\,p^{2}\left(1+4\frac{m^{2}}{p^{2}}-32\frac{m^{4}}{p^{4}}\right) \\
Im A_{4}(p^{2}) &=& \frac{1}{24} J_{0}\,p^{4}\left(1-8\frac{m^{2}}{p^{2}}+16\frac{m^{4}}{p^{4}}\right) \\
Im A_{5}(p^{2}) &=& -\frac{1}{48} J_{0}\,p^{4}\left(1+4\frac{m^{2}}{p^{2}}-32\frac{m^{4}}{p^{4}}\right). \label{Im A5}
\end{eqnarray}
Clearly, the imaginary parts (\ref{Im A1}) -- (\ref{Im A5}) of the amplitude
$T^{pv}_{\mu\nu\rho\sigma}$ satiesfy the WI (\ref{VWI1})--(\ref{VWI3}) with
$T_{i} \to Im A_{i}(p^{2})$, and the subtraction procedure we choose in the following
keeps this property for the entire formfactors $A_{i}(p^{2})$.\\

Now we start with an unsubtracted dispersion relation (DR) for the formfactors

\begin{equation}\label{unsubtracted DR}
T(p^{2}) = \frac{1}{\pi} \int\limits^{\infty}_{4m^{2}}\!\!\frac{dt}{t-p^{2}} Im T(t)
\end{equation}
and we observe that, for instance, the integral for $T_{1}(p^{2})$ is convergent
whereas for $T_{2}(p^{2})$ it is logarithmically divergent and needs to be subtracted
once, and for $T_{4}(p^{2})$ it is linearly divergent and needs to be subtracted
twice. We can infer already from the $p^{2} = t$ behaviour of the imaginary parts which
kind of dispersion relation we have to use, see Table~\ref{Subtraktionen}.

\begin{table}
\centering
\begin{minipage}{9.5cm}
\begin{tabular}{lr}
$T_{1}$,$T_{6}$,$A_{1}$ & unsubtracted dispersion relation\\ 
$T_{2}$,$T_{3}$,$T_{7}$,$T_{8}$,$A_{2}$,$A_{3}$ & once subtracted dispersion relation\\ 
$T_{4}$,$T_{5}$,$A_{4}$,$A_{5}$ & twice subtracted dispersion relation\\ 
\end{tabular}
\end{minipage}
\caption{The formfactors and the used type of dispersion relations}
\label{Subtraktionen}
\end{table}

So for the formfactors $T_{1}, T_{6}, A_{1}$ an unsubtracted DR is sufficient and
we get

\begin{eqnarray}
T_{1}(p^{2}) &=& \mp 4 T_{6}(p^{2}) = \frac{1}{2}A_{1}(p^{2}) {}\nonumber\\{}&=& -\frac{1}{4\pi} \int\limits^{\infty}_{4m^{2}}\!\!\frac{dt}{t-p^{2}} \frac{m^2}{t^2}\left(1-\frac{4m^{2}}{t}\right)^{ \hspace{-3pt}\frac{1}{2}} {}\nonumber\\{}&=& \frac{1}{p^{2}}\left[\frac{1}{24\pi}  - \frac{1}{2\pi} \frac{m^{2}}{p^{2}} + \frac{1}{2\pi} \frac{m^{2}}{p^{2}} a(p^{2})\right] \label{T1 in DR}
\end{eqnarray}
with

\begin{equation} \label{definition of a}
a(p^{2}) = \sqrt{\frac{4m^{2}-p^{2}}{p^{2}}} \arctan \sqrt{\frac{p^{2}}{4m^{2}-p^{2}}} \quad .
\end{equation}
A once subtracted DR defined by

\begin{equation}
T^{R}(p^{2}) = T(p^{2}) - T(0) = \frac{p^{2}}{\pi} \int\limits^{\infty}_{4m^{2}}\!\!\frac{dt}{t-p^{2}} \frac{1}{t}Im T(t)
\end{equation}
we use for the following formfactors (see Table 1)

\begin{eqnarray}
T^{R}_{2}(p^{2}) &=& \mp 4 T^{R}_{7}(p^{2}) = \frac{1}{2}A^{R}_{2}(p^{2}) {}\nonumber\\{}&=& \frac{p^{2}}{\pi} \int\limits^{\infty}_{4m^{2}}\!\!\frac{dt}{t-p^{2}} \frac{1}{t^{2}}\left(1-\frac{4m^{2}}{t}\right)^{\hspace{-3pt}-\frac{1}{2}} \hspace{-3pt} \left(- \frac{1}{48}t + \frac{1}{6}m^{2}-\frac{1}{3}\frac{m^{4}}{t}\right) {}\nonumber\\{}&=& -\frac{1}{18\pi} + \frac{1}{6\pi} \frac{m^{2}}{p^{2}} + \frac{1}{24\pi} \Bigl(1- 4 \frac{m^{2}}{p^{2}} \Bigr) a(p^{2})\\
T^{R}_{3}(p^{2}) &=& \frac{p^{2}}{\pi} \int\limits^{\infty}_{4m^{2}}\!\!\frac{dt}{t-p^{2}} \frac{1}{t^{2}}\left(1-\frac{4m^{2}}{t}\right)^{\hspace{-3pt}-\frac{1}{2}} \hspace{-3pt} \left( \frac{1}{96}t + \frac{1}{96}m^{2}-\frac{5}{24}\frac{m^{4}}{t}\right) {}\nonumber\\{}&=& \frac{7}{576\pi} + \frac{5}{48\pi} \frac{m^{2}}{p^{2}} - \frac{1}{48\pi} \Bigl( 1 + 5 \frac{m^{2}}{p^{2}} \Bigr) a(p^{2})\\
A^{R}_{3}(p^{2}) &=& \frac{p^{2}}{\pi} \int\limits^{\infty}_{4m^{2}}\!\!\frac{dt}{t-p^{2}} \frac{1}{t^{2}}\left(1-\frac{4m^{2}}{t}\right)^{\hspace{-3pt}-\frac{1}{2}} \hspace{-3pt} \left( \frac{1}{48}t + \frac{1}{12}m^{2}-\frac{2}{3}\frac{m^{4}}{t}\right) {}\nonumber\\{}&=& \frac{1}{72\pi} + \frac{1}{3\pi} \frac{m^{2}}{p^{2}} - \frac{1}{24\pi} \Bigl( 1 + 8 \frac{m^{2}}{p^{2}} \Bigr) a(p^{2})\\
T^{R}_{8}(p^{2}) &=& \mp \frac{p^{2}}{\pi} \int\limits^{\infty}_{4m^{2}}\!\!\frac{dt}{t-p^{2}} \frac{1}{t^{2}}\left(1-\frac{4m^{2}}{t}\right)^{\hspace{-3pt}-\frac{1}{2}} \hspace{-3pt} \left( \frac{1}{384}t + \frac{1}{96}m^{2}-\frac{1}{12}\frac{m^{4}}{t}\right) {}\nonumber\\{}&=&  \mp \frac{1}{576\pi} \mp \frac{1}{24\pi} \frac{m^{2}}{p^{2}} \pm \frac{1}{192\pi} \Bigl( 1 + 8 \frac{m^{2}}{p^{2}} \Bigr) a(p^{2}) \, .
\end{eqnarray}
For the remaining formfactors a twice subtracted DR defined by

\begin{equation}
T^{R}(p^{2}) = T(p^{2}) - T(0) -p^{2} \left. \frac{d}{dp^{2}}T(p^{2})\right|_{p^{2}=0} = \frac{p^{4}}{\pi} \int\limits^{\infty}_{4m^{2}}\!\!\frac{dt}{t-p^{2}} \frac{1}{t^{2}}Im T(t)
\end{equation}
is necessary and we find

\begin{eqnarray}
T^{R}_{4}(p^{2}) &=& \frac{1}{2}A^{R}_{4}(p^{2}) = \frac{p^{4}}{\pi} \int\limits^{\infty}_{4m^{2}}\!\!\frac{dt}{t-p^{2}} \frac{1}{t^{3}}\left(1-\frac{4m^{2}}{t}\right)^{\hspace{-3pt}-\frac{1}{2}} \hspace{-3pt} \left( \frac{1}{48}t^{2} - \frac{1}{6}t m^{2}+\frac{1}{3} m^{4} \right) {}\nonumber\\{}&=& p^{2}\left[\frac{1}{18\pi} - \frac{1}{6\pi}\frac{m^{2}}{p^{2}} - \frac{1}{24\pi} \Bigl(1-4\frac{m^{2}}{p^{2}} \Bigr) a(p^{2})\right]\\
T^{R}_{5}(p^{2}) &=& \frac{p^{4}}{\pi} \int\limits^{\infty}_{4m^{2}}\!\!\frac{dt}{t-p^{2}} \frac{1}{t^{3}}\left(1-\frac{4m^{2}}{t}\right)^{\hspace{-3pt}-\frac{1}{2}} \hspace{-3pt} \left( -\frac{1}{96}t^{2} + \frac{1}{48}t m^{2}+\frac{1}{12} m^{4} \right) {}\nonumber\\{}&=& p^{2}\left[-\frac{5}{288\pi} - \frac{1}{24\pi} \frac{m^{2}}{p^{2}} + \frac{1}{48\pi} \Bigl(1+ 2 \frac{m^{2}}{p^{2}} \Bigr) a(p^{2})\right]\\
A^{R}_{5}(p^{2}) &=& \frac{p^{4}}{\pi} \int\limits^{\infty}_{4m^{2}}\!\!\frac{dt}{t-p^{2}} \frac{1}{t^{3}}\left(1-\frac{4m^{2}}{t}\right)^{\hspace{-3pt}-\frac{1}{2}} \hspace{-3pt} \left( -\frac{1}{48}t^{2} - \frac{1}{12}t m^{2}+\frac{2}{3}m^{4}\right) {}\nonumber\\{}&=& p^{2}\left[-\frac{1}{72\pi} - \frac{1}{3\pi} \frac{m^{2}}{p^{2}} + \frac{1}{24\pi} \Bigl(1+ 8 \frac{m^{2}}{p^{2}} \Bigr) a(p^{2})\right]. \label{A5R in DR}
\end{eqnarray}

With these explicit expressions for the formfactors we have determined the whole
amplitude $T_{\mu\nu\rho\sigma}$,
Eqs.(\ref{amplitude expressed by T-product})--(\ref{formfactors3}), from which the
correct Ward identities will follow.

\section{'t Hooft--Veltman regularization}

Although the method of dispersion relations appears quite different to the methods
of regularizations we can reproduce its results in a definite regularization scheme,
namely in the n-dimensional regularization procedure of \, 't Hooft--Veltman
\cite{HooftVeltman}. It is instructive to work it out in more detail.

We start with amplitude
(\ref{amplitude by Dimensional regularization with massive fermions}), calculate the
$\gamma$-matrices and follow the standard procedure by inserting the Feynman parameter
integral
\begin{equation}
\frac{1}{a b} = \int\limits_{0}^{1} \!dx \ \frac{1}{[a x + b(1-x)]^{2}} \, \, .
\end{equation}
Then we obtain for the `no interchange' amplitudes of the pure
tensor pieces

\begin{eqnarray}
&&T^{ni}_{\mu \nu \rho \sigma}(p) = - i \frac{2^{\omega}}{32} \int\limits_{0}^{1} \!dx \int\!\! \frac{d^{2\omega}l}{(2\pi)^{2\omega}} \frac{(2l-p(1-2x))_{\nu} (2l-p(1-2x))_{\sigma}}{[l^{2}-\Delta]^{2}} {}\nonumber\\{}&&\quad \times \Bigl[ (l+px)_{\mu}(l-p(1-x))_{\rho} + (l+px)_{\rho}(l-p(1-x))_{\mu} - g_{\mu\rho} (l+px)^{\lambda}(l-p(1-x))_{\lambda} \Bigr]
\end{eqnarray}
and

\begin{equation}
T^{dv,ni}_{\mu \nu \rho \sigma}(p) = - i \frac{2^{\omega}}{32} m^{2} g_{\mu\rho} \int\limits_{0}^{1} \!dx \int\!\! \frac{d^{2\omega}l}{(2\pi)^{2\omega}} \frac{(2l-p(1-2x))_{\nu} (2l-p(1-2x))_{\sigma}}{[l^{2}-\Delta]^{2}} \, \, ,
\end{equation}
with

\begin{equation}
\Delta := m^{2} - p^{2}(1-x) + p^{2}(1-x)^{2} = m^{2} - p^{2}x(1-x) \, .
\end{equation}
Calculating the 't Hooft--Veltman integrals
\begin{eqnarray}
P_{0} &=& \int\!\! \frac{d^{2\omega}l}{(2\pi)^{2\omega}} \frac{1}{[l^{2}-\Delta]^{\alpha}} = \frac{(-1)^{\alpha}i} {(4\pi)^{\omega}} \frac{\Gamma(\alpha - \omega)}{\Gamma(\alpha)} \Delta^{\omega-\alpha} \\
P_{1}^{\mu\nu} &=& \int\!\! \frac{d^{2\omega}l}{(2\pi)^{2\omega}} \frac{l^{\mu}l^{\nu}}{[l^{2}-\Delta]^{\alpha}} = \frac{\Delta}{2(\omega - \alpha + 1)} g^{\mu\nu} P_{0} \label{integral P1}\\
P_{2}^{\mu\nu\rho\sigma} &=& \int\!\! \frac{d^{2\omega}l}{(2\pi)^{2\omega}} \frac{l^{\mu} l^{\nu} l^{\rho} l^{\sigma}} {[l^{2}-\Delta]^{\alpha}} {}\nonumber\\{}&=& \frac{\Delta^{2}}{4(\omega - \alpha + 1)(\omega - \alpha + 2)} (g^{\mu\nu} g^{\rho\sigma} + g^{\mu\rho} g^{\nu\sigma} + g^{\mu\sigma} g^{\nu\rho}) P_{0} \label{integral P2}
\end{eqnarray}
with $\alpha=2$ provides the following expressions
\begin{eqnarray} 
&&T^{ni}_{\mu \nu \rho \sigma}(p) = - i \frac{2^{\omega}}{32} \int\limits_{0}^{1} \!dx P_{0} \Biggl\{ -2x(1-x)(1-2x)^{2} p_{\mu}p_{\nu}p_{\rho}p_{\sigma} + (1-2x)^{2} \frac{\Delta}{\omega-1} {}\nonumber\\{}&&\qquad\times (p_{\mu}p_{\nu} g_{\rho \sigma} + p_{\mu}p_{\sigma} g_{\nu \rho} + p_{\nu}p_{\rho} g_{\mu \sigma} + p_{\rho}p_{\sigma} g_{\mu \nu}) - 4x(1-x)\frac{\Delta}{\omega-1} p_{\mu}p_{\rho} g_{\nu \sigma} {}\nonumber\\{}&&\qquad + \left[ -(1+\omega)(1-2x)^{2} \frac{\Delta}{\omega-1} + x(1-x)(1-2x)^{2}p^{2} \right] p_{\nu}p_{\sigma} g_{\mu \rho} {}\nonumber\\{}&& \qquad + 2 \frac{\Delta^{2}} {\omega(\omega-1)} (g_{\mu \nu} g_{\rho \sigma} + g_{\mu \sigma} g_{\nu \rho}) + \left[ -2\omega \frac{\Delta^{2}} {\omega(\omega-1)} +2x(1-x)p^{2} \frac{\Delta}{\omega-1} \right] g_{\mu \rho} g_{\nu \sigma} \Biggr\}\label{explicit Tni in massive dim reg}\\ 
&&T^{dv,ni}_{\mu \nu \rho \sigma}(p) = - i \frac{2^{\omega}}{32} m^{2} g_{\mu\rho} \int\limits_{0}^{1} \!dx P_{0} \Biggl\{ 2 \frac{\Delta}{\omega-1} g_{\nu\sigma} + (1-2x)^{2} p_{\nu}p_{\sigma} \Biggr\}\label{explicit Tdvni in massive dim reg}.
\end{eqnarray}
These expressions we compare now with the formfactor decompositions (\ref{formfactors2})
and (\ref{formfactors3}) (recall Eqs.(\ref{without interchange}) and (\ref{axial part with
gamma5 symmetrically distributed})) then we obtain all formfactors of the amplitude
$T_{\mu\nu\rho\sigma}$ explicitly
\begin{eqnarray}
T_{1}(p^{2}) &=& \mp 4\ T_{6}(p^{2})=i \frac{2^{\omega}}{4} \int\limits^{1}_{0}\!dx \ x(1-x)(1-2x)^{2} P_{0} \label{T1 in massive dim reg} \\
T_{2}(p^{2}) &=& \mp 4\ T_{7}(p^{2})=- i \frac{2^{\omega}}{8} \int\limits^{1}_{0}\!dx (1-2x)^{2} \frac{\Delta}{\omega-1} P_{0} \\
T_{3}(p^{2}) &=&  = - i \frac{2^{\omega}}{32} \int\limits^{1}_{0}\!dx P_{0} \Biggl\{ - 4x(1-x) \frac{\Delta}{\omega-1} + (2p^{2}x(1-x)-m^{2})(1-2x)^{2} \Biggr\} \\
T_{4}(p^{2}) &=& - i \frac{2^{\omega}}{4} \int\limits^{1}_{0}\!dx \frac{\Delta^{2}}{\omega(\omega-1)} P_{0} \label{T4 in massive dim reg}\\
T_{5}(p^{2}) &=& - i \frac{2^{\omega}}{8} \int\limits^{1}_{0}\!dx P_{0} \Biggl\{ p^{2}x(1-x) \frac{\Delta}{\omega-1} - \frac{\Delta^{2}}{\omega} \Biggr\} \label{T5 in massive dim reg}\\
T_{8}(p^{2}) &=& \pm i \frac{2^{\omega}}{64} \int\limits^{1}_{0}\!dx (8x^{2}-8x+1) \frac{\Delta}{\omega-1} P_{0} \, .
\end{eqnarray}
The formfactors of the amplitude $T^{pv}_{\mu\nu\rho\sigma}$ follow via
Eq.(\ref{pure vector amplitude})
\begin{eqnarray}
A_{1}(p^{2}) &=& 2\ T_{1}(p^{2}) = i \frac{2^{\omega}}{2} \int\limits^{1}_{0}\!dx \ x(1-x)(1-2x)^{2} P_{0} \label{A1 in massive dim reg} \\
A_{2}(p^{2}) &=& 2\ T_{2}(p^{2}) = - i \frac{2^{\omega}}{4} \int\limits^{1}_{0}\!dx (1-2x)^{2} \frac{\Delta}{\omega-1} P_{0} \\
A_{3}(p^{2}) &=& - i \frac{2^{\omega}}{16} \int\limits^{1}_{0}\!dx P_{0} \Biggl\{ - 4x(1-x) \frac{\Delta}{\omega-1} + 2p^{2}x(1-x)(1-2x)^{2} \Biggl\} \\
A_{4}(p^{2}) &=& 2\ T_{4}(p^{2}) = - i \frac{2^{\omega}}{2} \int\limits^{1}_{0}\!dx \frac{\Delta^{2}}{\omega(\omega-1)} P_{0} \\
A_{5}(p^{2}) &=& - i \frac{2^{\omega}}{4} \int\limits^{1}_{0}\!dx P_{0} \Biggl\{ \bigl( m^{2}+p^{2}x(1-x)\bigr) \frac{\Delta}{\omega-1} - \frac{\Delta^{2}}{\omega} \Biggl\}. \label{A5 in massive dim reg}
\end{eqnarray}
As we can see the formfactors $T_{1}, T_{6}$ and $A_{1}$ (Eqs.(\ref{T1 in massive dim reg})and
(\ref{A1 in massive dim reg})) are finite in the limit $\omega \to 1$ and explicitly we find
\begin{eqnarray}
T_{1}(p^{2}) &=& \mp 4 T_{6}(p^{2}) = \frac{1}{2}A_{1}(p^{2}) = -\frac{1}{8\pi} \int\limits^{1}_{0}\!dx \frac{x(1-x)(1-2x)^{2}}{m^{2}-p^{2}x(1-x)} {}\nonumber\\{}&=& \frac{1}{p^{2}}\left[\frac{1}{24\pi} - \frac{1}{2\pi} \frac{m^{2}}{p^{2}} + \frac{1}{2\pi} \frac{m^{2}}{p^{2}} a(p^{2})\right] \label{T1R in massive dim reg}
\end{eqnarray}
with $a(p^{2})$ given by Eq.(\ref{definition of a}).

However, the remaining formfactors are divergent in the limit $\omega \to 1$ and we
have to renormalize them in an appropriate way. We separate the formfactors into a
divergent part and a finite part
\begin{equation} \label{pole terms}
T_{i} = \frac{1}{1 - \omega}T^{pol}_{i} + T^{fin}_{i} \, ,
\end{equation}
so that we can extract the finite result by a suitable prescription. The formula we
need for this procedure is
\begin{eqnarray}
\frac{\Gamma(1-\omega)}{(2\pi)^{\omega}} \Delta^{\omega-1}f(\omega) &=& \frac{f(1)}{2\pi} \Bigl[  \frac{1}{1-\omega} - \ln \frac{\Delta}{2\pi} - \gamma \Bigr] - \frac{1}{2\pi}\left.\frac{df}{d\omega}\right|_{\omega=1}+ O(1-\omega) \, .
\end{eqnarray}
It is the pole part $T^{pol}_{i}$ of the formfactor which tells us which kind of
renormalization we have to choose, for instance, for a constant a simple subtraction
is sufficient, for a term proportional to $p^{2}$ a double subtraction.

In this way the following formfactors are determined by a simple subtraction
\begin{eqnarray}
T^{R}_{2}(p^{2}) &=& \mp 4 T^{R}_{7} (p^{2}) = \frac{1}{2}A^{R}_{2} (p^{2}) = T_{2}(p^{2}) - T_{2}(0) {}\nonumber\\{}&=& \frac{1}{16\pi} \int\limits^{1}_{0}\!dx (1-2x)^{2} \ln \frac{m^{2}-p^{2}x(1-x)}{m^{2}} {}\nonumber\\{}&=& -\frac{1}{18\pi} + \frac{1}{6\pi} \frac{m^{2}}{p^{2}} + \frac{1}{24\pi} \Bigl(1- 4 \frac{m^{2}}{p^{2}}\Bigr) a(p^{2})  \label{T2R in massive dim reg}\\
T^{R}_{3}(p^{2}) &=& T_{3}(p^{2}) - T_{3}(0) = -\frac{1}{16\pi} \int\limits^{1}_{0}\!dx \ x(1-x)  \ln \frac{m^{2}-p^{2}x(1-x)}{m^{2}} {}\nonumber\\{}&& + \frac{1}{64\pi} \int\limits^{1}_{0}dx (1-2x)^{2} \frac{p^{2}x(1-x)}{m^{2}-p^{2}x(1-x)} {}\nonumber\\{}&=& \frac{7}{576\pi} + \frac{5}{48\pi} \frac{m^{2}}{p^{2}} - \frac{1}{48\pi} \Bigl( 1 + 5 \frac{m^{2}}{p^{2}} \Bigr) a(p^{2}) \\
A^{R}_{3}(p^{2}) &=& A_{3}(p^{2}) - A_{3}(0) = -\frac{1}{8\pi} \int\limits^{1}_{0}\!dx \ x(1-x)  \ln \frac{m^{2}-p^{2}x(1-x)}{m^{2}} {}\nonumber\\{}&& + \frac{1}{16\pi} \int\limits^{1}_{0}dx (1-2x)^{2} \frac{p^{2}x(1-x)}{m^{2}-p^{2}x(1-x)} {}\nonumber\\{}&=& \frac{1}{72\pi} + \frac{1}{3\pi} \frac{m^{2}}{p^{2}} - \frac{1}{24\pi} \Bigl( 1 + 8 \frac{m^{2}}{p^{2}} \Bigr) a(p^{2}) \label{A3R in massive dim reg}\\
T^{R}_{8}(p^{2}) &=& T_{8}(p^{2}) - T_{8}(0) = \mp \frac{1}{128\pi} \int\limits^{1}_{0}\!dx  \Bigl( 8x^{2}-8x+1 \Bigr)  \ln \frac{m^{2}-p^{2}x(1-x)}{m^{2}} {}\nonumber\\{}&=& \mp \frac{1}{576\pi} \mp \frac{1}{24\pi} \frac{m^{2}}{p^{2}} \pm \frac{1}{192\pi} \Bigl( 1 + 8 \frac{m^{2}}{p^{2}} \Bigr) a(p^{2}) \, ,
\end{eqnarray}
whereas for the remaining formfactors a double subtraction is needed
\begin{eqnarray} 
T^{R}_{4}(p^{2}) &=& \frac{1}{2} A^{R}_{4}(p^{2}) = T_{4}(p^{2}) - T_{4}(0) - p^{2} \left.\frac{d}{dp^{2}}T_{4}(p^{2})\right|_{p^{2}=0} {}\nonumber\\{} &=& \frac{1}{8\pi} \int\limits^{1}_{0}\!dx \Biggl[ \Bigl( m^{2}-p^{2}x(1-x) \Bigr) \ln \frac{m^{2}-p^{2}x(1-x)}{m^{2}} + p^{2}x(1-x) \Biggr] {}\nonumber\\{} &=& p^{2}\left[ \frac{1}{18\pi} - \frac{1}{6\pi} \frac{m^{2}}{p^{2}} - \frac{1}{24\pi} \Bigl(1-4 \frac{m^{2}}{p^{2}} \Bigr) a(p^{2})\right]\label{T4R in massive dim reg}\\
T^{R}_{5}(p^{2}) &=&  T_{5}(p^{2}) - T_{5}(0) - p^{2} \left.\frac{d}{dp^{2}}T_{5}(p^{2})\right|_{p^{2}=0} {}\nonumber\\{} &=& \frac{1}{16\pi} \int\limits^{1}_{0}\!dx \ p^{2}x(1-x) \ln \frac{m^{2}-p^{2}x(1-x)}{m^{2}} {}\nonumber\\{}&=& p^{2}\left[-\frac{5}{288\pi} - \frac{1}{24\pi} \frac{m^{2}}{p^{2}} + \frac{1}{48\pi} \Bigl(1+ 2 \frac{m^{2}}{p^{2}} \Bigr) a(p^{2})\right]\\
A^{R}_{5}(p^{2}) &=&  A_{5}(p^{2}) - A_{5}(0) - p^{2} \left.\frac{d}{dp^{2}}A_{5}(p^{2})\right|_{p^{2}=0} {}\nonumber\\{} &=& \frac{1}{8\pi} \int\limits^{1}_{0}\!dx \Biggl[\Bigl( m^{2} + p^{2}x(1-x) \Bigr) \ln \frac{m^{2}-p^{2}x(1-x)}{m^{2}} + p^{2}x(1-x)\Biggr] {}\nonumber\\{}&=& p^{2}\left[-\frac{1}{72\pi} - \frac{1}{3\pi} \frac{m^{2}}{p^{2}} + \frac{1}{24\pi} \Bigl(1+8 \frac{m^{2}}{p^{2}} \Bigr) a(p^{2})\right]. \label{A5R in massive dim reg}
\end{eqnarray}
Of course, the formfactors $A^{R}_{i}$ of the tensor amplitude $T^{pv}_{\mu\nu\rho\sigma}$
satisfy the WI (\ref{VWI1}) -- (\ref{VWI3}) and as the difference
$T^{dv}_{\mu\nu\rho\sigma}$ to the pure tensor amplitude is proportional to $m^{2}$
(recall Eqs.(\ref{pure vector amplitude}),(\ref{explicit Tdvni in massive dim reg})) the
amplitude $T^{V}_{\mu\nu\rho\sigma}$ will also satisfy the WI (\ref{VWI1}) -- (\ref{VWI3})
in the limit $m \to 0$ .

This procedure works in complete analogy to the dispersion relation approach (see Table~1)
and we clearly obtain the same results for the formfactors as one can see by comparing
Eqs.(\ref{T1R in massive dim reg})--(\ref{A5R in massive dim reg}) with 
(\ref{T1 in DR})--(\ref{A5R in DR}). Even more, as we shall show in the next chapter,
the two approaches are equivalent.

But before we want to consider the case of massless fermions, the limit $m \to 0$,
in order to compare our results with the ones of other authors. Taking the limit
$m \to 0$ in the $1/(1 - \omega)$ expansion (Eq.(\ref{pole terms})) of the formfactors
we find the following results
\begin{eqnarray}
T_{1}(p^{2}) &=& \mp 4\ T_{6}(p^{2}) = \frac{1}{24\pi p^{2}} \label{T1 for WI}\\
T_{2}(p^{2}) &=& -\frac{1}{p^{2}}T_{4}(p^{2}) = \mp 4\ T_{7}(p^{2}) {}\nonumber\\{}&=& \frac{1}{48\pi} \Biggl[ \frac{1}{\omega-1} + \gamma - \frac{8}{3} + \ln \left|\frac{p^{2}}{2\pi\mu^{2}}\right| - i\pi \Biggr]  \label{T2 for WI} \\
T_{3}(p^{2}) &=& -\frac{1}{p^{2}}T_{5}(p^{2}) = \mp 4\ T_{8}(p^{2}) {}\nonumber\\{}&=& -\frac{1}{96\pi} \Biggl[ \frac{1}{\omega-1} + \gamma - \frac{2}{3} + \ln\left|\frac{p^{2}}{2\pi\mu^{2}}\right| - i\pi \Biggr]  \label{T3 for WI}.
\end{eqnarray}
which agree with the expressions we obtain when starting with $m = 0$ from the very
beginning (we also introduced here the mass $\mu$ to keep the correct dimensionality,
see Ref.\cite{Kohlprath}).

Now, for comparison Tomiya \cite{Tomiya} in his work on the gravitational anomaly
defines the amplitude with the covariant $T^{\star}$-product and calculates the following
formfactors
\begin{eqnarray}
T^{T}_{1}(p^{2})&=&\mp 4\ T^{T}_{6}(p^{2}) = \frac{1}{24\pi p^{2}} \label{T1 from Tomiya}\\
T^{T}_{2}(p^{2})&=&-\frac{1}{p^{2}}T^{T}_{4}(p^{2})=\mp 4\ T^{T}_{7}(p^{2}) = 0 \label{T2 from Tomiya}\\
T^{T}_{3}(p^{2})&=&-\frac{1}{p^{2}}T^{T}_{5}(p^{2})=\mp 4\ T^{T}_{8}(p^{2}) = -\frac{1}{48\pi} \, \, . \label{T3 from Tomiya}
\end{eqnarray}
They also satiesfy the WI (\ref{VWI1}) -- (\ref{VWI3}), however, they (some of them) differ
from ours (\ref{T1 for WI}) -- (\ref{T3 for WI}), which is not surprising as the covariant
$T^{\star}$-product differs from the ordinary $T$-product by suitable seagull terms.
But as we shall show below his and our results lead to the same amplitudes and consequently
to the same anomaly expression.

Using on one hand the WI (\ref{VWI1}) -- (\ref{VWI3}) for the formfactors $T_{i}(p^{2})$
in the massless limit, $m \to 0$, and on the other the fact that
\begin{equation} \label{axial--vector}
T_{6}=\mp \frac{1}{4}T_{1} \, ,\qquad T_{7}=\mp\frac{1}{4}T_{2} \, , \qquad T_{8}=\mp\frac{1}{4}T_{3} \, ,
\end{equation}
we find for the several amplitudes (\ref{formfactors1}) -- (\ref{formfactors3}) the
following expressions
\begin{eqnarray}
T_{0000} &=& \left(\mp p_{0}p_{1}^{3} + p_{1}^{4}\right)T_{1}(p^{2}) \\
T_{0001} &=& \frac{1}{4}\left(4 p_{0}p_{1}^{3}\mp 3 p_{0}^{2}p_{1}^{2}\mp p_{1}^{4}\right)T_{1}(p^{2}) \\
T_{0011} &=& T_{0101}=\frac{1}{2}\left(\mp p_{0}^{3}p_{1} + 2p_{0}^{2}p_{1}^{2}\mp p_{0}p_{1}^{3}\right)T_{1}(p^{2}) \\
T_{0111} &=& \frac{1}{4}\left(\mp p_{0}^{4} + 4p_{0}^{3}p_{1}\mp 3p_{0}^{2}p_{1}^{2}\right)T_{1}(p^{2}) \\
T_{1111} &=& \left(p_{0}^{4}\mp p_{0}^{3}p_{1}\right)T_{1}(p^{2}) \, . 
\end{eqnarray}
As we can see all amplitudes depend only on the (convergent) formfactor $T_{1}(p^{2})$,
explicitly given by Eq.(\ref{T1 for WI}), and are independent of a second (divergent)
formfactor, say $T_{2}(p^{2})$, which one might have expected to contribute too (since we have
six restrictive relations between the eight formfactors $T_{i}(p^{2})$).

So the amplitudes do not determine the formfactors uniquely! Therefore $T_{2}(p^{2})$ can be
chosen at will and Tomiya \cite{Tomiya} in his work renormalized the formfactors conveniently
such that $T_{2}(p^{2})$ came out to be zero. This is in accordance with the amplitude
result of Deser and Schwimmer \cite{DeserSchwimmer}, \cite{Deser} who consider fields
without chirality.

On the other hand, Alvarez-Gaum\'e and Witten \cite{AlvarezWitten} consider the 
gravitational anomaly within light-cone coordinates and use a specific regularization
prescription to calculate the following amplitude
\begin{equation}
U(p) = i\int\!d^{2}x\ e^{ipx}\langle 0 \vert T \lbrack T_{++}(x)T_{++}(0) \rbrack \vert 0 \rangle =\frac{1}{24\pi}\frac{p_{+}^{3}}{p_{-}}=\frac{1}{12\pi p^{2}}p_{+}^{4} \, \, .
\end{equation}
Their result is consistent with our calculations as we have
\begin{eqnarray}
T_{++++} &=&\frac{1}{4}\left( T_{0000} + 4 T_{0001} + 2 T_{0011} + 4 T_{0101} + 4 T_{0111} + T_{1111} \right) {}\nonumber\\{}&=&\frac{1}{4}\frac{1}{24\pi p^{2}}\Bigl[ (1\mp1)p_{0}^{4}+(4\mp4)p_{0}^{3}p_{1}+(6\mp6)p_{0}^{2}p_{1}^{2}+(4\mp4)p_{0}p_{1}^{3} +(1\mp1)p_{1}^{4}\Bigr] {}\nonumber\\{}&=& \frac{1\mp1}{24\pi p^{2}}p_{+}^{4} \, \, .
\end{eqnarray}

\section{Equivalence of dispersion relations and dimensional regularization}

The dimensional regularization procedure is a method which satisfies the WI
(\ref{VWI1}) -- (\ref{VWI3}). On the other hand, within the dispersion relation approach
we also have effectively renormalized in a way which keeps this WI property
(\ref{VWI1}) -- (\ref{VWI3}). If we had subtracted more often or at some other point than
$p^{2}=0$ we would have lost this property.

The dispersive approach and the dimensional regularization procedure are equivalent in the
following sense. For the formfactors that were convergent or logarithmically divergent the
corresponding integrals in both approaches do not have just identical
values but we even can transform them into each other by a suitable substitution. More
precise, we transform the Feynman parameter integrals over x in the dimensional
regularization procedure into the corresponding dispersion integrals over t. 
For the formfactors that were linearly divergent the situation is as follows. Accidentially for 
$T_{5}$ there exists such a substitution, however, not for $T_{4} = \frac{1}{2}A_{4}$ and
$A_{5}$. But anyway, for these formfactors the integrals are identical, what is actually
sufficient.

The substitutions which link the two approaches are
\begin{equation} \label{substitution dis rel - dim reg}
y = 1-2x \, \, , \qquad t = \frac{4m^{2}}{1-y^{2}} \, \, .
\end{equation}
To show the equivalence we need the following relations which are valid for any function $f$
with suitable differentiation properties
\begin{eqnarray}
\frac{1}{\pi} \int\limits^{\infty}_{4m^{2}} \!\!\frac{dt}{t\left(t-p^{2}\right)} \left(1-\frac{4m^{2}}{t}\right)^{\!-1/2} \!\!\!f(t) &=& \frac{1}{2\pi} \int\limits^{1}_{0}\!dx \frac{f\left(\frac{m^{2}}{x(1-x)}\right)}{m^{2}-p^{2}x(1-x)}\\
\int\limits^{1}_{0} \!dx \ f'(x) \ln \frac{m^{2}-p^{2}x(1-x)}{m^{2}} &=& -\int\limits^{1}_{0}\!dx \ f(x) \frac{p^{2}(-1+2x)}{m^{2}-p^{2}x(1-x)} \quad .
\end{eqnarray}

\section{Anomalous Ward identities and gravitational anomalies}

Now we turn to the calculation of the Ward identities and gravitational anomalies. We consider
the massless limit, $m \to 0$, where the formfactors $2T_{i} \to A_{i}$ (i = 1,...,5) fulfill
the WI (\ref{VWI1}) -- (\ref{VWI3}). This means that the WI for the pure tensor part
(\ref{pTV}) is satiesfied
\begin{equation}
p^{\mu}T^{V}_{\mu\nu\rho\sigma}(p)=0 \, .
\end{equation}
Calculating next the WI for the pseudo tensor part (\ref{pTA}) we use the formfactor identities
(\ref{axial--vector}) and we find
\begin{eqnarray}
p^{\mu} T^{A}_{\mu \nu \rho \sigma}(p) &=& \pm \frac{1}{4}T_{2}\ \varepsilon_{\nu\tau}p^{\tau}(p_{\rho}p_{\sigma}-g_{\rho\sigma}p^{2}) {}\nonumber\\{}&&\pm \frac{1}{4}T_{3} \left( \varepsilon_{\rho\tau}p^{\tau}(p_{\nu}p_{\sigma}-g_{\nu\sigma}p^{2}) + \varepsilon_{\sigma\tau}p^{\tau}(p_{\nu}p_{\rho}-g_{\nu\rho}p^{2}) \right).
\end{eqnarray}
In the flat space-time limit $g_{\mu\nu}(x) = \eta_{\mu\nu}$ we have the relation
\begin{equation}
\varepsilon_{\nu\tau}p^{\tau}(p_{\rho}p_{\sigma}-g_{\rho\sigma}p^{2}) =\varepsilon_{\rho\tau}p^{\tau}(p_{\nu}p_{\sigma}-g_{\nu\sigma}p^{2}) = \varepsilon_{\sigma\tau}p^{\tau}(p_{\nu}p_{\rho}-g_{\nu\rho}p^{2})
\end{equation}
and we finally obtain the anomalous result
\begin{equation} 
p^{\mu} T^{A}_{\mu \nu \rho \sigma}(p) = \mp \frac{1}{4} p^{2}T_{1}\ \varepsilon_{\nu\tau}p^{\tau}(p_{\rho}p_{\sigma}-g_{\rho\sigma}p^{2}) \, .
\end{equation}
As we discovered already before when calculating the amplitudes, the anomalous WI depends
only on the finite formfactor $T_{1} = \mp 4 T_{6}$ with its explicit result (\ref{T1 for WI}).
So the anomaly is independent of a specific renormalization procedure (as long as it preserves
the WI (\ref{VWI1}) -- (\ref{VWI3})) and we clearly agree with the
anomaly results of Tomiya \cite{Tomiya} and Alvarez-Gaum\'e and Witten \cite{AlvarezWitten}.

We want to emphasize that our subtraction procedure is {\em the\/} `natural' choice dictated
by the $t-$behaviour of the imaginary parts $ImT_{i}(t)$ of the formfactors. Since on general
grounds the imaginary parts of the amplitudes fulfill the WI (\ref{VWI1}) -- (\ref{VWI3})
(in the limit $m \to 0$) the 'naturally' chosen dispersion relations
\begin{eqnarray}
&&\frac{1}{\pi}\int\limits^{\infty}_{4m^{2}}\!\!\frac{dt}{t-p^{2}}\frac{p^{2}}{t}\left[t\, Im\,T_{1}(t)+Im\,T_{2}(t)+2\, Im\,T_{3}(t)\right]=0 \\
&&\frac{1}{\pi}\int\limits^{\infty}_{4m^{2}}\!\!\frac{dt}{t-p^{2}}\frac{p^{4}}{t^2}\left[t\, Im\,T_{2}(t)+Im\,T_{4}(t)\right]=0 \\
&&\frac{1}{\pi}\int\limits^{\infty}_{4m^{2}}\!\!\frac{dt}{t-p^{2}}\frac{p^{4}}{t^2}\left[t\, Im\,T_{3}(t)+Im\,T_{5}(t)\right]=0
\end{eqnarray}
imply the pure tensor WI (\ref{VWI1}) -- (\ref{VWI3}) for the renormalized formfactors
(in the limit $m \to 0$)
\begin{eqnarray}
&&p^{2}T_{1}(p^{2})+T_{2}^{R}(p^{2})+2\,T_{3}^{R}(p^{2})=0\\
&&p^{2}T_{2}^{R}(p^{2})+T_{4}^{R}(p^{2})=0\\
&&p^{2}T_{3}^{R}(p^{2})+T_{5}^{R}(p^{2})=0 \, .
\end{eqnarray}
Therefore our subtraction procedure automatically shifts the total anomaly into the
pseudotensor part of the WI (\ref{pTA}).\\

What is the origin of the anomaly in this dispersive approach? The source of the anomaly
is the existence of a superconvergence sum rule for the imaginary part of the formfactor
$T_{1}(p^{2})$
\begin{equation}
\int\limits^{\infty}_{0}\!dt\,Im\,T_{1}(t)=-\frac{m^2}{4}\int\limits^{\infty}_{4m^{2}}\!\!\frac{dt}{t^{2}}\left(1-\frac{4m^{2}}{t}\right)^{\hspace{-3pt}\frac{1}{2}}=-\frac{1}{24} \, \, .
\end{equation} 
The anomaly originates from a $\delta$-function singularity of $ImT_{1}(t)$ when the
threshold $t = 4m^{2} \to 0$ approaches zero (the infrared region)
\begin{equation}
\lim_{m\rightarrow 0} Im\,T_{1}(t)=-\lim_{m\rightarrow 0}\frac{m^{2}}{4t^{2}}\left(1-\frac{4m^{2}}{t}\right)^{\hspace{-3pt}\frac{1}{2}}\theta(t-4m^{2})=-\frac{1}{24}\delta(t) \, .
\end{equation}
The limit must be performed in a distributional sense.\\
Then the unsubtracted dispersion relation for $T_{1}(p^{2})$, Eq.(\ref{unsubtracted DR}), 
provides the result (\ref{T1 for WI}).
This threshold singularity of the imaginary part of the relevant formfactor is a typical
feature of the dispersion relation approach for calculating the anomaly
(see e.g. Refs.\cite{Bertlmann}, \cite{Horejsi} -- \cite{HorejsiSchnabl}, and the
Appendix B in Ref. \cite{Deser}).\\

Next we turn to the energy-momentum tensor. From the anomalous WI
\begin{equation} \label{anomalous VWI consistend with all results}
p^{\mu} T_{\mu \nu \rho \sigma}(p) = \mp \frac{1}{96\pi} \varepsilon_{\nu\tau}p^{\tau}(p_{\rho}p_{\sigma}-g_{\rho\sigma}p^{2})
\end{equation}
we can deduce the linearized consistent Einstein (or diffeomorphism) anomaly
\begin{equation} \label{linearized consistent Einstein anomaly}
\partial^{\mu}\langle T_{\mu\nu} \rangle = \mp \frac{1}{192\pi} \varepsilon_{\mu\nu} \partial^{\mu} \left( \partial_{\alpha}\partial_{\beta} h^{\alpha\beta} - \partial_{\alpha}\partial^{\alpha}h^{\beta}_{\ \beta}\right).
\end{equation}
For comparison we demonstrate that result (\ref{linearized consistent Einstein anomaly}) is
indeed the linearization of the exact result that follows from differential geometry and
topology (see for instance Ref.\cite{Bertlmann}). The exact Einstein anomaly in two dimensions
is given by
\begin{eqnarray}
G^{E}\left(v_{\xi},\Gamma \right) &=& -\int\!d^{2}x\ e\xi_{\nu}\nabla_{\mu}\langle T^{\mu\nu}\rangle = \mp\frac{1}{96\pi} \int\limits_{M_{2}}tr\ v_{\xi}d\Gamma = \pm\frac{1}{96\pi} \int\!d^{2}x\ \varepsilon^{\gamma\delta}\xi^{\beta}\partial_{\alpha}\partial_{\gamma}\Gamma^{\alpha}_{\ \delta\beta} \label{consistent Einstein anomaly in two dimensions},
\end{eqnarray}
where
\begin{equation}
\left(v_{\xi}\right)^{\beta}_{\ \alpha} = \partial_{\alpha}\xi^{\beta} \, .
\end{equation}
Therefore we get
\begin{equation} \label{exact consistent Einstein anomaly}
\nabla^{\mu}\langle T_{\mu\nu}\rangle = \mp\frac{1}{96\pi}\frac{1}{e}\varepsilon^{\gamma\delta}\partial_{\alpha}\partial_{\gamma}\Gamma^{\alpha}_{\ \delta\nu} \, \, .
\end{equation}
Considering the linearized gravitational field, Eq.(\ref{linearization of the vielbein with
upper indices}), the Christoffel symbols become (with explicit $\kappa$-dependence)
\begin{eqnarray} \label{linearization of the Christoffel symbols}
\Gamma^{\alpha}_{\ \nu\beta}&=&\frac{1}{2}g^{\alpha\lambda}\left(\partial_{\beta} g_{\lambda\nu} - \partial_{\lambda} g_{\nu\beta} + \partial_{\nu} g_{\beta\lambda}\right) = \frac{\kappa}{2} \left( \partial_{\beta}h^{\alpha}_{\ \nu}-\partial^{\alpha}h_{\nu\beta}+\partial_{\nu}h_{\beta}^{\ \alpha}\right) + O\left(\kappa^{2}\right), 
\end{eqnarray}
so that we find as linearization of the exact result (\ref{exact consistent Einstein anomaly})
\begin{equation}
\partial^{\mu}\langle T_{\mu\nu}\rangle = \mp\frac{1}{192\pi} \varepsilon^{\gamma\delta}\partial_{\alpha}\partial_{\gamma}\left( \partial_{\nu}h^{\alpha}_{\ \delta}-\partial^{\alpha}h_{\delta\nu}\right).
\end{equation}
This agrees with our result (\ref{linearized consistent Einstein anomaly}) since in two
dimensions we have the identity
\begin{equation}\label{epsilon identity}
\varepsilon_{\mu\beta}\partial^{\mu}\left(\partial_{\alpha}\partial_{\nu}h^{\alpha\beta}-\partial_{\alpha}\partial^{\alpha}h_{\nu}^{\ \beta}\right) = \varepsilon_{\mu\nu}\partial^{\mu}\left(\partial_{\alpha}\partial_{\beta}h^{\alpha\beta}-\partial_{\alpha}\partial^{\alpha}h_{\beta}^{\ \beta}\right).
\end{equation}

Now what about the covariant Einstein anomaly? It arises when considering the covariantly
transforming energy-momentum tensor $\tilde T_{\mu\nu}$ which is related to our tensor
definition (\ref{energy-momentum-tensor}) by the Bardeen-Zumino polynomial
$\mathcal{P}_{\mu\nu}$ \cite{BardeenZumino}
\begin{equation}
\langle \tilde T_{\mu\nu}\rangle = \langle T_{\mu\nu}\rangle + \mathcal{P}_{\mu\nu} \, .
\end{equation}
This polynomial is calculable and explicitly we find (in two dimensions) \cite{Bertlmann}
\begin{eqnarray}
\nabla^{\mu}\mathcal{P}_{\mu\nu}&=&\mp\frac{1}{96\pi}\frac{1}{e}\left(\varepsilon_{\mu\nu}\nabla^{\mu}\mathcal{R}- \varepsilon^{\gamma\delta}\partial_{\alpha}\partial_{\gamma}\Gamma^{\alpha}_{\ \delta\nu}\right)
\end{eqnarray}
($\mathcal{R}$ denotes the Ricci scalar) which leads to the covariant Einstein anomaly
\begin{eqnarray}
\nabla^{\mu}\langle \tilde T_{\mu\nu}\rangle &=&\mp\frac{1}{96\pi}\frac{1}{e}\varepsilon_{\mu\nu}\nabla^{\mu}\mathcal{R} \, .
\end{eqnarray}
Linearizing the Ricci scalar (with explicit $\kappa$-dependence)
\begin{equation} 
\mathcal{R}=\kappa\left( \partial_{\mu}\partial_{\nu}h^{\mu\nu}-\partial_{\mu}\partial^{\mu}h^{\nu}_{\ \nu}\right) + O\left(\kappa^{2}\right)\label{linearization of R}
\end{equation}
and using (\ref{linearization of the Christoffel symbols}), (\ref{epsilon identity}) we get
\begin{equation}
\partial^{\mu}\mathcal{P}_{\mu\nu} = \mp \frac{1}{192\pi} \varepsilon_{\mu\nu} \partial^{\mu} \left( \partial_{\alpha}\partial_{\beta} h^{\alpha\beta} - \partial_{\alpha}\partial^{\alpha}h^{\beta}_{\ \beta}\right)
\end{equation}
so that we find for the linearized covariant Einstein anomaly
\begin{equation}
\partial^{\mu}\langle\tilde T_{\mu\nu} \rangle = \mp \frac{1}{96\pi} \varepsilon_{\mu\nu} \partial^{\mu} \left( \partial_{\alpha}\partial_{\beta} h^{\alpha\beta} - \partial_{\alpha}\partial^{\alpha}h^{\beta}_{\ \beta}\right).
\end{equation}
It is twice the linearized consistent result (\ref{linearized consistent Einstein anomaly})
as it should be.\\

Finally we also calculate the trace identity, Eqs.(\ref{gTV}) and (\ref{gTA}). Using again
relations (\ref{axial--vector}) we find
\begin{equation}
T^{\mu}_{\ \mu\rho\sigma}=\left(p^{2}T_{1}+2\omega T_{2}+4 T_{3}\right)\Biggl[\left(p_{\rho}p_{\sigma}-p^{2}g_{\rho\sigma}\right) \mp\frac{1}{4}\left( \varepsilon_{\rho\lambda}p^{\lambda}p_{\sigma}+\varepsilon_{\sigma\lambda}p^{\lambda}p_{\rho}\right)\Biggl],
\end{equation}
and taking into account the WI (\ref{VWI1}) provides us the anomalous result
\begin{equation}
T^{\mu}_{\ \mu\rho\sigma}=-p^{2}T_{1}\Biggl[\left(p_{\rho}p_{\sigma}-p^{2}g_{\rho\sigma}\right) \mp\frac{1}{4}\left( \varepsilon_{\rho\lambda}p^{\lambda}p_{\sigma}+\varepsilon_{\sigma\lambda}p^{\lambda}p_{\rho}\right)\Biggl].
\end{equation}
The anomalous TI depends only on the finite formfactor $T_{1} = \mp 4 T_{6}$
-- what we know already from the previous discussion --
so that it is independent of a specific renormalization procedure (as long as it preserves 
the WI (\ref{VWI1}) -- (\ref{VWI3})).

Inserting the formfactor, Eq.(\ref{T1 for WI}), gives
\begin{equation} \label{anomalous TI consistent with all results}
T^{\mu}_{\ \mu\rho\sigma}=-\frac{1}{24\pi} \Biggl[\left(p_{\rho}p_{\sigma}-p^{2}g_{\rho\sigma}\right) \mp\frac{1}{4}\left( \varepsilon_{\rho\lambda}p^{\lambda}p_{\sigma}+\varepsilon_{\sigma\lambda}p^{\lambda}p_{\rho}\right)\Biggl],
\end{equation}
which implies the following linearization of the Weyl (or trace) anomaly
\begin{equation} \label{linearized Weyl anomaly}
\langle T^{\mu}_{\ \mu} \rangle = \frac{1}{48\pi}\Biggl[ \left(\partial_{\mu}\partial_{\nu}h^{\mu\nu}-\partial_{\mu}\partial^{\mu}h^{\nu}_{\ \nu}\right) \mp\frac{1}{2}\varepsilon_{\mu\lambda}\partial^{\lambda}\partial_{\nu}h^{\mu\nu}\Biggr].
\end{equation}

Again, we compare our calculation with the exact result for the Weyl anomaly which is
given by (see for instance Ref.\cite{Bertlmann})
\begin{equation} \label{consistent Weyl anomaly in two dimensions}
G^{W}(\sigma) = \int\!d^{2}x\ e\ \sigma \langle T^{\mu}_{\ \mu} \rangle = \frac{1}{48\pi} \int\!d^{2}x\ e\ \sigma\left( \mathcal{R} \mp \frac{1}{2} \varepsilon^{ab} \nabla_{\mu}\omega_{ab}^{\ \ \mu}\right)
\end{equation}
and consequently we have
\begin{equation} \label{exact consistent Weyl anomaly}
\langle T^{\mu}_{\ \mu}\rangle = \frac{1}{48\pi} \left( \mathcal{R} \mp \frac{1}{2} \varepsilon^{ab} \nabla_{\mu}\omega_{ab}^{\ \ \mu}\right).
\end{equation}
Considering the linearizations of the spin connection (with explicit $\kappa$-dependence)
\begin{equation}
\omega^{a}_{\ b\mu}=e^{a}_{\ \nu}\nabla_{\mu}E_{b}^{\ \nu} = e^{a}_{\ \nu}\partial_{\mu}E_{b}^{\ \nu}+ e^{a}_{\ \nu}\Gamma^{\nu}_{\ \mu\lambda} E_{b}^{\ \lambda}= \frac{\kappa}{2}\left( \partial_{b}h_{\mu}^{\ a}-\partial^{a}h_{b\mu}\right) + O\left(\kappa^{2}\right) \label{linearization of omega}
\end{equation}
and of the Ricci scalar Eq.(\ref{linearization of R}), we see that 
our result (\ref{linearized Weyl anomaly}) is indeed the linearization of
the exact expression (\ref{exact consistent Weyl anomaly}).

Adding last but not least the Bardeen-Zumino polynomial $\mathcal{P}_{\mu\nu}$ 
\begin{equation}
\mathcal{P}^{\mu}_{\ \mu}=\pm \frac{1}{96\pi}\varepsilon^{ab}\nabla^{\mu}\omega_{ab\mu}
\end{equation}
with its linearization
\begin{equation}
\mathcal{P}^{\mu}_{\ \mu}=\pm \frac{1}{96\pi}\varepsilon^{ab}\partial^{\mu}\partial_{b}h_{\mu a}
\end{equation}
we find for the covariant trace anomaly
\begin{equation}
\langle T^{\mu}_{\ \mu}\rangle = \frac{1}{48\pi}\mathcal{R}
\end{equation}
and
\begin{equation}
\langle T^{\mu}_{\ \mu} \rangle = \frac{1}{48\pi}\left( \partial_{\mu}\partial_{\nu}h^{\mu\nu}-\partial_{\mu}\partial^{\mu}h^{\nu}_{\ \nu}\right)
\end{equation}
for its linearized version. Clearly these results are in agreement with Ref.
\cite{Leutwyler}.

\section{Schwinger terms}

In quantum field theory the energy-momentum tensors form an algebra which is generally not
closed but has central extensions, socalled Schwinger terms, for example
\begin{eqnarray}
\lbrack T_{00}(x), T_{00}(0) \rbrack_{ET} &=& i \left(T_{01}(x) + T_{01}(0) \right) \partial_{1}\delta(x^{1}) + S_{0000}\\ 
\lbrack T_{01}(x), T_{01}(0) \rbrack_{ET} &=& i \left(T_{01}(x) + T_{01}(0) \right) \partial_{1}\delta(x^{1}) + S_{0101}\\ 
\lbrack T_{00}(x), T_{01}(0) \rbrack_{ET} &=& i \left(T_{00}(x) + T_{00}(0) \right) \partial_{1}\delta(x^{1}) + S_{0001} \, . 
\end{eqnarray}
The Schwinger terms, the c-number terms $S_{0000}, S_{0101}, S_{0001}$, can be determined by
considering the vacuum expectation value of the ETC.

To evaluate the ST we work with  a technique that has been introduced by K\"all\'en
\cite{Kaellen} and is closely related to the dispersive approach used before. This technique
has been applied already by S\'ykora \cite{Sykora} to compute the ST for currents in Yang-Mills
theories. Our aim is to generalize this procedure to the case of gravitation, where the current
is replaced by the energy-momentum tensor.

From the Lagrangian (\ref{Lagrangian}) describing a Weyl fermion in a gravitational background
field in two dimensions we get the following classical energy-momentum tensor
\begin{equation}
T_{\mu\nu}=\ :\frac{i}{4}\bar\psi\left(\gamma_{\mu} \stackrel{\leftrightarrow}{\partial}_{\nu}+\gamma_{\nu} \stackrel{\leftrightarrow}{\partial}_{\mu}\right) P_{\pm}\psi:\ =\frac{1}{2}\left( T_{\mu\nu}^{V}\pm T_{\mu\nu}^{A}\right),
\end{equation}
where $::$ means normal ordering. Using relation (\ref{elimination of gamma5}) and the
equations of motions we can express the pseudo tensor part of the enery-momentum tensor by the
pure tensor part (recall that the tensor is symmetric)
\begin{equation}
T_{\mu\nu}^{A}= -\varepsilon_{\mu}^{\ \lambda}T_{\lambda\nu}^{V} \, \, .
\end{equation}
In two dimensions we have the identity
\begin{equation}
\varepsilon_{\mu}^{\ \lambda}\varepsilon_{\rho}^{\ \tau}=-g_{\mu\rho}g^{\lambda\tau}+g^{\lambda}_{\ \rho}g^{\tau}_{\ \mu} \, \, ,
\end{equation}
so that we find
\begin{eqnarray} \label{symbolic VEV of commutator}
&&\langle 0\vert [T_{\mu\nu}(x),T_{\rho\sigma}(0)]\vert 0\rangle = \frac{1}{4} \Biggl\{ \langle 0\vert [T^{V}_{\mu\nu}(x),T^{V}_{\rho\sigma}(0)]\vert 0\rangle + \langle 0\vert [T^{V}_{\rho\nu}(x),T^{V}_{\mu\sigma}(0)]\vert 0\rangle {}\nonumber\\{}&& \qquad- g_{\mu\rho} \langle 0\vert [T^{V\lambda}_{\nu}(x),T^{V}_{\lambda\sigma}(0)]\vert 0\rangle  \mp \varepsilon_{\mu}^{\ \lambda}\langle 0\vert [T^{V}_{\lambda\nu}(x),T^{V}_{\rho\sigma}(0)]\vert 0\rangle \mp \varepsilon_{\rho}^{\ \lambda} \langle 0\vert [T^{V}_{\mu\nu}(x),T^{V}_{\lambda\sigma}(0)]\vert 0\rangle \Biggr\}.
\end{eqnarray}
Let us define
\begin{equation}
F_{\mu\nu\rho\sigma}(x):= \langle 0\vert T^{V}_{\mu\nu}(x) T^{V}_{\rho\sigma}(0) \vert 0\rangle.
\end{equation}
By inserting the completeness relations $\sum_{n}\vert n \rangle\langle n\vert=\mathbf{1}$ and
using the translation invariance
\begin{equation}
T^{V}_{\mu\nu}(x)=e^{iPx}T^{V}_{\mu\nu}(0)e^{-iPx} \, ,
\end{equation}
we obtain
\begin{eqnarray}
F_{\mu\nu\rho\sigma}(x)&=& \sum_{n}\langle 0\vert T_{\mu\nu}^{V}(x) \vert n\rangle\langle n\vert T_{\rho\sigma}^{V}(0)\vert 0\rangle {}\nonumber\\{}
&=& \sum_{n}\langle 0\vert T_{\mu\nu}^{V}(0) \vert n\rangle\langle n\vert T_{\rho\sigma}^{V}(0)\vert 0\rangle e^{-ip_{n}x} \, ,
\end{eqnarray}
where the sum runs over many-particle states $\vert n\rangle$ with positive energy and momentum
$p_{n}$.\\
We may write
\begin{equation}
F_{\mu\nu\rho\sigma}(x) = \int\!d^{2}p\ e^{-ipx}G_{\mu\nu\rho\sigma}(p)\theta(p^{0}) \, ,
\end{equation}
where
\begin{equation} \label{symbolic Gmunurhosigma}
G_{\mu\nu\rho\sigma}(p)=\sum_{n}\delta(p_{n}-p)\langle 0\vert T_{\mu\nu}^{V}(0) \vert n\rangle\langle n\vert T_{\rho\sigma}^{V}(0)\vert 0\rangle.
\end{equation}
From Lorentz covariance and symmetry we get the same decompositon into formfactors as for
$T^{V}_{\mu\nu\rho\sigma}$ (see Eq. (\ref{formfactors2})
\begin{eqnarray}
G_{\mu \nu \rho \sigma}(p) &=& p_{\mu}p_{\nu}p_{\rho}p_{\sigma} G_{1}(p^{2}) + (p_{\mu}p_{\nu} g_{\rho \sigma} + p_{\rho}p_{\sigma} g_{\mu \nu}) G_{2}(p^{2}) {} \nonumber \\ {} & &+ (p_{\mu}p_{\rho} g_{\nu \sigma} + p_{\mu}p_{\sigma} g_{\nu \rho} + p_{\nu}p_{\rho} g_{\mu \sigma} + p_{\nu}p_{\sigma} g_{\mu \rho}) G_{3}(p^{2}) {} \nonumber \\ {} & &+ g_{\mu \nu}g_{\rho \sigma} G_{4}(p^{2}) + (g_{\mu \rho}g_{\nu \sigma} + g_{\mu \sigma}g_{\nu \rho}) G_{5}(p^{2}) \, .
\end{eqnarray}
Making use of $\partial^{\mu} T^{V}_{\mu\nu}(x)=0$ provides the Ward identity
$p^{\mu}G_{\mu\nu\rho\sigma}(p)=0$ that can be expressed by the formfactors
\begin{eqnarray}
p^{2}G_{1}+G_{2}+2G_{3}&=&0 \label{VWI1 for G}\\
p^{2}G_{2}+G_{4}&=&0 \label{VWI2 for G}\\
p^{2}G_{3}+G_{5}&=&0 \label{VWI3 for G} \,.
\end{eqnarray}
Now let us explicitly evaluate $G_{\mu\nu\rho\sigma}(p)$. As we are considering the 
energy-momentum tensor as a free (non interacting) tensor -- analogously to the case of free
currents -- we only need to sum up states that consist of one fermion-antifermion pair in
Eq. (\ref{symbolic Gmunurhosigma}). We get
\begin{equation}
G_{\mu\nu\rho\sigma}(p)=\int\!dp_{1}\int\!dp_{2}\sum_{s_{1}}\sum_{s_{2}}\ \delta(p-p_{1}-p_{2})\langle 0\vert T_{\mu\nu}^{V}(0) \vert n\rangle\langle n\vert T_{\rho\sigma}^{V}(0)\vert 0\rangle,
\end{equation}
where
\begin{equation}
\vert n\rangle = b^{\dagger(s_{1})}(p_{1}) d^{\dagger(s_{2})}(p_{2})\vert 0\rangle.
\end{equation}
Let us assume that the fermion fields are canonically quantized
\begin{eqnarray}
\psi(x)&=& \frac{1}{(2\pi)^{1/2}}\int\!dp\sum_{s=1}^{2}\sqrt{\frac{m}{E_{p}}}\left[ b^{(s)}(p)u^{(s)}(p)e^{-ipx}+d^{\dagger(s)}(p)v^{(s)}(p)e^{ipx}\right]\\
\bar\psi(x)&=& \frac{1}{(2\pi)^{1/2}}\int\!dp\sum_{s=1}^{2}\sqrt{\frac{m}{E_{p}}}\left[ b^{\dagger(s)}(p)\bar u^{(s)}(p)e^{ipx}+d^{(s)}(p)\bar v^{(s)}(p)e^{-ipx}\right],
\end{eqnarray}
then we find
\begin{eqnarray}
\langle 0\vert T_{\mu\nu}^{V}(0)\vert n\rangle &=& \frac{1}{8\pi}\frac{m}{\sqrt{E_{p_{1}}E_{p_{2}}}}\bar v^{(s_{2})}(p_{2})\left(\gamma_{\mu}(p_{1}-p_{2})_{\nu}+\gamma_{\nu}(p_{1}-p_{2})_{\mu}\right) u^{(s_{1})}(p_{1})\\
\langle n\vert T_{\rho\sigma}^{V}(0)\vert 0\rangle &=& -\frac{1}{8\pi}\frac{m}{\sqrt{E_{p_{1}}E_{p_{2}}}}\bar u^{(s_{1})}(p_{1})\left(\gamma_{\rho}(p_{1}-p_{2})_{\sigma}+\gamma_{\sigma}(p_{1}-p_{2})_{\rho}\right) v^{(s_{2})}(p_{2}) \, .
\end{eqnarray}
This provides
\begin{eqnarray}
G_{\mu\nu\rho\sigma}(p) &=&-\frac{1}{64 \pi^{2}} \int\!dp_{1}\int\!dp_{2}\sum_{s_{1}}\sum_{s_{2}}\ \delta(p-p_{1}-p_{2})\frac{m^{2}}{E_{p_{1}}E_{p_{2}}}{}\nonumber\\{}&&\times  (p_{1}-p_{2})_{\nu}(p_{1}-p_{2})_{\sigma}\bar v^{(s_{2})}(p_{2})\gamma_{\mu}u^{(s_{1})}(p_{1})\bar u^{(s_{1})}(p_{1})\gamma_{\rho}v^{(s_{2})}(p_{2})  {}\nonumber\\{}&&+ (\mu \leftrightarrow \nu) + (\rho \leftrightarrow \sigma) + {\mu \leftrightarrow \nu \choose \rho \leftrightarrow \sigma}.
\end{eqnarray}
Without the interchanges we call this $G^{ni}_{\mu\nu\rho\sigma}(p)$.\\
If we use the completeness relations
\begin{eqnarray}
\sum_{s=1}^{2}u^{(s)}(p)\bar u^{(s)}(p) &=& \frac{/\hspace{-6.5pt}p+m}{2m} \\
\sum_{s=1}^{2}v^{(s)}(p)\bar v^{(s)}(p) &=& \frac{/\hspace{-6.5pt}p-m}{2m}
\end{eqnarray}
and the integral equation
\begin{equation}
\int\!\!\frac{dp}{2E_{p}}f(p) = \int\!d^{2}p\ \delta(p^{2}-m^{2}) \theta(p^{0}) f(p),
\end{equation}
we obtain
\begin{eqnarray}
G^{ni}_{\mu\nu\rho\sigma}&=&-\frac{1}{64 \pi^{2}}\int\!d^{2}p_{1}d^{2}p_{2}\ \delta(p-p_{1}-p_{2})\delta(p_{1}^{2}-m^{2})\delta(p_{2}^{2}-m^{2})\theta(p_{1}^{0})\theta(p_{2}^{0}) {}\nonumber\\{}&&\times (p_{1}-p_{2})_{\nu}(p_{1}-p_{2})_{\sigma}tr\Bigl[ \gamma_{\mu}(/\hspace{-6.5pt}p_{1}+m)\gamma_{\rho}(/\hspace{-6.5pt}p_{2}-m) \Bigr].
\end{eqnarray}
Integrating next over the first $\delta$-function and evaluating the trace gives
\begin{eqnarray}
G^{ni}_{\mu\nu\rho\sigma}&=&-\frac{1}{32 \pi^{2}}\int\!d^{2}p_{1}\ \delta(p_{1}^{2}-m^{2})\delta((p-p_{1})^{2}-m^{2})\theta(p_{1}^{0})\theta(p^{0}-p_{1}^{0}) {}\nonumber\\{}&&\times (2p_{1}-p)_{\nu}(2p_{1}-p)_{\sigma}\Bigl[ p_{1\mu}(p-p_{1})_{\rho}+p_{1\rho}(p-p_{1})_{\mu} {}\nonumber\\{}&& -g_{\mu\rho}p_{1}^{\lambda}(p-p_{1})_{\lambda}-g_{\mu\rho}m^{2} \Bigr].
\end{eqnarray}
If we compare this with (\ref{abstract Im Tni}) and (\ref{abstract Im Tdvni}) ($p_{1}=-k$),
we see that $G_{\mu\nu\rho\sigma}(p)$ is given by the imaginary part of the amplitude of the
pure vector loop
\begin{equation} \label{G and Tpv}
G_{\mu\nu\rho\sigma}(p)=\frac{1}{2\pi^{2}} Im T^{pv}_{\mu\nu\rho\sigma}(p) \, .
\end{equation}
This is the important relation which links the Schwinger terms to the gravitational anomalies.

From Eq. (\ref{Im A1}) -- (\ref{Im A5}) we read off the formfactors
\begin{eqnarray}
G_{1}(p^{2}) &=& -\frac{1}{4\pi^{2}} J_{0}\frac{m^{2}}{p^{2}}\left(1 -4\frac{m^{2}}{p^{2}}\right) \label{explicit G1}\\
G_{2}(p^{2}) &=& -\frac{1}{48\pi^{2}} J_{0}p^{2}\left(1-8\frac{m^{2}}{p^{2}}+16\frac{m^{4}}{p^{4}}\right) \\
G_{3}(p^{2}) &=& \frac{1}{96\pi^{2}} J_{0}p^{2}\left(1+4\frac{m^{2}}{p^{2}}-32\frac{m^{4}}{p^{4}}\right) \\
G_{4}(p^{2}) &=& \frac{1}{48\pi^{2}} J_{0}p^{4}\left(1-8\frac{m^{2}}{p^{2}}+16\frac{m^{4}}{p^{4}}\right) \\
G_{5}(p^{2}) &=& -\frac{1}{96\pi^{2}} J_{0}p^{4}\left(1+4\frac{m^{2}}{p^{2}}-32\frac{m^{4}}{p^{4}}\right). \label{explicit G5}
\end{eqnarray}
Now we consider the commutator
\begin{equation} \label{[T,T] in terms of G}
\langle 0\vert [T^{V}_{\mu\nu}(x), T^{V}_{\rho\sigma}(0)] \vert 0\rangle = F_{\mu\nu\rho\sigma}(x)-F_{\rho\sigma\mu\nu}(-x) = \int\!d^{2}p\ e^{-ipx} \varepsilon(p^{0}) G_{\mu\nu\rho\sigma}(p).
\end{equation}
If we remove the mass, $m \rightarrow 0$, that acted as an infrared cutoff, we get
\begin{equation}
G_{1}(p^{2}) = \lim_{m\rightarrow 0}-\frac{1}{4\pi^{2}}\frac{m^2}{p^4}\left(1-\frac{4m^{2}}{p^{2}}\right)^{\hspace{-3pt}\frac{1}{2}}\theta\left(p^{2}-4m^{2}\right) = -\frac{1}{24\pi^{2}}\delta(p^{2}) \, .
\end{equation}
From Eq. (\ref{[T,T] in terms of G}) we explicitly find
\begin{eqnarray}
\langle 0\vert [T^{V}_{00}(x), T^{V}_{00}(0)] \vert 0\rangle_{ET} &=& \lim_{x_{0}\rightarrow 0} -\frac{1}{24\pi^{2}}\int\!d^{2}p\ e^{-ipx}p_{1}^{4}\ \varepsilon(p^{0}) \delta(p^{2}) = 0 \\
\langle 0\vert [T^{V}_{11}(x), T^{V}_{11}(0)] \vert 0\rangle_{ET} &=& \lim_{x_{0}\rightarrow 0} -\frac{1}{24\pi^{2}}\int\!d^{2}p\ e^{-ipx}p_{0}^{4}\ \varepsilon(p^{0}) \delta(p^{2}) = 0 \\
\langle 0\vert [T^{V}_{00}(x), T^{V}_{11}(0)] \vert 0\rangle_{ET} &=& \langle 0\vert [T^{V}_{01}(x), T^{V}_{01}(0)] \vert 0\rangle_{ET} {}\nonumber\\{}&=&  \lim_{x_{0}\rightarrow 0} -\frac{1}{24\pi^{2}}\int\!d^{2}p\ e^{-ipx}p_{0}^{2}p_{1}^{2}\ \varepsilon(p^{0}) \delta(p^{2}) = 0 \\
\langle 0\vert [T^{V}_{00}(x), T^{V}_{01}(0)] \vert 0\rangle_{ET} &=& \lim_{x_{0}\rightarrow 0} -\frac{1}{24\pi^{2}}\int\!d^{2}p\ e^{-ipx}p_{0}p_{1}^{3}\ \varepsilon(p^{0}) \delta(p^{2}) \\
\langle 0\vert [T^{V}_{01}(x), T^{V}_{11}(0)] \vert 0\rangle_{ET} &=& \lim_{x_{0}\rightarrow 0} -\frac{1}{24\pi^{2}}\int\!d^{2}p\ e^{-ipx}p_{0}^{3}p_{1}\ \varepsilon(p^{0}) \delta(p^{2}) \, .\label{[T01,T11] no interchange} 
\end{eqnarray}
The first three expressions vanish because $\varepsilon(p^{0})$ is antisymmetric. 
To evaluate the next two we use the Pauli-Jordan function 
\begin{equation}
\triangle(x)=\frac{1}{2\pi} \int\!d^{2}p\ e^{-ipx} \varepsilon(p^{0})\delta(p^{2})
\end{equation}
with the properties
\begin{eqnarray}
\partial^{\mu}\partial_{\mu} \triangle(x)&=&0\\
\left. \triangle(x)\right|_{x^{0}=0}&=&0\\
\left. \partial_{0}\triangle(x)\right|_{x^{0}=0}&=&-i\delta(x^{1})
\end{eqnarray}
and we find
\begin{equation}
\lim_{x_{0}\rightarrow 0}-\frac{1}{24\pi^{2}}\int\!d^{2}p\ e^{-ipx}\varepsilon(p^{0}) p_{0}p_{1}p^{2}\delta(p^{2}) = \lim_{x_{0}\rightarrow 0}-\frac{1}{12\pi}\partial_{0}\partial_{1}(\partial_{0}^{2}-\partial_{1}^{2})\triangle(x)=0 \, .
\end{equation}
So we conclude
\begin{eqnarray}
\langle 0\vert [T^{V}_{00}(x), T^{V}_{01}(0)] \vert 0\rangle_{ET} &=& \langle 0\vert [T^{V}_{01}(x), T^{V}_{11}(0)] \vert 0\rangle_{ET} {}\nonumber\\{}&=& \lim_{x_{0}\rightarrow 0}-\frac{1}{12\pi}\partial_{1}^{3}\partial_{0}\triangle(x) = \frac{i}{12\pi}\partial_{1}^{3}\delta(x^{1}) \, .
\end{eqnarray}
With relation (\ref{symbolic VEV of commutator}) we finally obtain the Schwinger terms in the
ETC of the energy-momentum tensors
\begin{eqnarray}
\langle 0\vert [T_{00}(x), T_{00}(0)] \vert 0\rangle_{ET} &=& \langle 0\vert [T_{11}(x), T_{11}(0)] \vert 0\rangle_{ET} =\mp\frac{i}{24\pi}(\partial_{1})^{3}\delta(x^{1})\\ 
\langle 0\vert [T_{00}(x), T_{11}(0)] \vert 0\rangle_{ET} &=& \langle 0\vert [T_{01}(x), T_{01}(0)] \vert 0\rangle_{ET} =\mp\frac{i}{24\pi}(\partial_{1})^{3}\delta(x^{1})\\
\langle 0\vert [T_{00}(x), T_{01}(0)] \vert 0\rangle_{ET} &=& \langle 0\vert [T_{01}(x), T_{11}(0)] \vert 0\rangle_{ET} =\frac{i}{24\pi}(\partial_{1})^{3}\delta(x^{1}) \, .
\end{eqnarray}
Our result agrees with the one of Tomiya \cite{Tomiya} who works with an invariant spectral
function method and in addition uses a cohomological approach. It also coincides with the
result of Ebner, Heid and Lopes-Cardoso \cite{EbnerHeidLopes} who derive the Schwinger terms
directly from the gravitational anomaly. In our approach Eq. (\ref{G and Tpv}) is the basic
relation. It connects the  Schwinger terms, determined by $G_{\mu\nu\rho\sigma}$, with the
(linearized) gravitational anomalies given by $Im\,T^{pv}_{\mu\nu\rho\sigma}$ in our dispersion
relations procedure.

\section{Conclusions}

We have investigated the gravitational anomalies, specifically the pure Einstein anomaly and
the Weyl anomaly. So we demanded the quantized energy-momentum tensor to be symmetric -- no
Lorentz anomaly occurs -- which is a possible choice. The relevant amplitude, the two-point
function of the energy-momentum tensors $T_{\mu\nu\rho\sigma}$, we have separated into its
pure tensor part $T^{V}_{\mu\nu\rho\sigma}$ and into its pseudo-tensor part
$T^{A}_{\mu\nu\rho\sigma}$,
Eqs.(\ref{amplitude expressed by T-product}) -- (\ref{formfactors3}), 
and we have decomposed the amplitudes into a general structure of tensors containing
8 formfactors $T_{1}(p^{2}), ... ,T_{8}(p^{2})$. 

These formfactors we have expressed by
dispersion relations where we had to calculate only the imaginary parts via the Cutkosky
rules. Our subtraction procedure for the formfactors
-- no subtraction for $T_{1}, T_{6}$, one subtraction for $T_{2}, T_{3}, T_{7}, T_{8}$
and two subtractions for $T_{4}, T_{5}$ -- is {\em the\/} 'natural' choice dictated
by the $t-$behaviour of the imaginary parts $ImT_{i}(t)$. It implies that the pure tensor WI
(\ref{VWI1}) -- (\ref{VWI3}) for the renormalized formfactors is satisfied (in the limit
$m \to 0$), so that the total anomaly is automatically shifted into the
pseudotensor part of the WI (\ref{pTA}). It turns out that the anomalous Ward identity and
the anomalous trace identity depend only on the finite formfactor $T_{1} = \mp 4 T_{6}$,
with its explicit result (\ref{T1 for WI}), demonstrating such the independence of a special
renormalization procedure. From the anomalous Ward identity and the anomalous trace identity
we could deduce the linearized gravitational anomalies
-- the linearized Einstein- and Weyl anomaly -- and determine their covariant versions.

The origin of the anomaly is the existence of a superconvergence sum rule for the imaginary
part of the formfactor $T_{1}(p^{2})$. In the zero mass limit the imaginary part
of the formfactor approaches a $\delta$-function singularity at zero momentum squared,
exhibiting in this way the infrared feature of the gravitational anomalies.
This is an independent and complementary view of the anomalies as compared to the ultraviolet
regularization procedures. If we compare, however, the DR approach with the n-dimensional
regularization procedure of \, 't~Hooft--Veltman we find an equivalence. The two approaches
are linked by the substitutions (\ref{substitution dis rel - dim reg}).

We have also calculated the gravitational Schwinger terms which occur in the ETC of the
energy-momentum tensors. We have adapted our dispersive approach to the method of K\"all\'en.
As a result all gravitational Schwinger terms are determined by the formfactor $G_{1}(p^{2})$
which in the zero mass limit approaches a $\delta$-function singularity at zero momentum
squared -- as in the case of anomalies. So also the Schwinger terms show this peculiar
infrared feature of the anomalies.

We have performed all calculations in two dimensions, where already all essential features of
the DR approach show up, analogously to the chiral current case. From a practical point of
view the method appears quite appealing. All one has to calculate is the imaginary part of
an amplitude (formfactor), which is an easy task. However, this computational simplicity
is a special (and convenient) feature of the two space-time dimensions. In higher dimensions
the amplitude will contain more formfactors and we have to calculate dispersion relations for
higher loop diagrams, which is a much more delicate task, but nevertheless we expect the
method to work here too.


\end{document}